\documentclass[12pt,preprint]{aastex}
\usepackage{psfig}

\shorttitle{H$_2$ excitation in ULIRGs}
\shortauthors{}

\begin{document}

\title{Molecular Hydrogen Excitation in Ultraluminous Infrared
Galaxies\footnote{Based on observations at the European Southern
Observatory VLT (67.B-0417).}}

\author{R.I. Davies}
\affil{Max-Planck-Institut f\"ur extraterrestrische Physik, 
Postfach 1312, 85741, Garching, Germany}
%\email{davies@mpe.mpg.de}

\author{A. Sternberg}
\affil{School of Physics and Astronomy, Tel Aviv University, Tel Aviv
69978, Israel}

\author{M. Lehnert}
\affil{Max-Planck-Institut f\"ur extraterrestrische Physik, 
Postfach 1312, 85741, Garching, Germany}

\and

\author{L.E. Tacconi-Garman}
\affil{European Southern Observatory, Karl Schwarzschildstrasse 2,
85748 Garching, Germany}

\begin{abstract}
We report medium resolution VLT ISAAC K-band spectroscopy of the
nuclei of seven ultraluminous infrared galaxies.
After accounting for stellar absorption features, we have detected
several molecular hydrogen (H$_2$) $v=$1-0, 2-1, and 3-2
vibrational emission lines, as well as the H{\small I} Br$\gamma$ and
He{\small \,I} $2^1P-2^1S$ recombination lines.
The relative H$_2$ line intensities show little variation between the
objects, suggesting that the H$_2$ excitation mechanisms in the nuclei are
similar in all the objects.
The 1-0 emissions appear thermalised at temperatures
$T\sim1000$\,K.
However, the 2-1 and 3-2 emissions show evidence of being
radiatively excited by far-ultraviolet (FUV) photons, suggesting that
the H$_2$ excitation in the ULIRGs may arise in dense photon dominated
regions (PDRs).
We show that the line ratios in the nuclei are consistent with PDRs
with cloud densities between $10^4$ to 10$^5$\,cm$^{-3}$, 
exposed to far ultraviolet (FUV)
radiation fields at least $10^3$ times more intense than the
ambient FUV intensity in the local interstellar medium.
We have constructed starburst models for the ULIRGs based on their
H$_2$ properties, as well as on the intensities of the 
recombination lines. Our models provide a consistent picture
of young 1-5\,Myr star clusters surrounded by relatively dense PDRs which are
irradiated by intense FUV fluxes.
Comparison to the inner few hundred parsecs of the Milky Way indicates
that the star formation efficiency in ULIRGs is 10--100 times higher
than in the Galactic Center.
%The equivalent widths and ratio of the 1-0\,S(1) and Br$\gamma$ lines
%can change away from the bright continuum nucleus,
%suggesting that the circumnuclear excitation mechanisms may differ
%between objects.
\end{abstract}


\section{Introduction}
\label{sec:intro}

In star forming galaxies, molecular hydrogen (H$_2$)
vibrational emission lines include some of the brightest
lines in the near-infrared (1.9-2.5 $\mu$m) 
K-band, and have therefore been candidates for detailed studies.
The H$_2$ lines may be used to probe the physical conditions
of molecular clouds near the star forming regions.
For example, the relative
strengths of lines in the $v=1-0$ and $v=2-1$ bands allow 
radiative fluorescent excitation in low-density
photodissociation regions (PDRs)
\citep[]{bla87,ste88,ste89,bur90b,dra96}
to be distinguished from
collisional excitation in shocks or dense PDRs
\citep[]{ste89,bur90a,cha91,dra96,lim02}.
Determining the H$_2$ excitation mechanism
in a given source generally requires the observation of
many, often weak lines, particularly
since different processes (e.g.~shocks vs.~FUV heating) can
give rise to similar intense and thermalised 1-0 emissions.
This requirement
has resulted in a great deal of uncertainty in the
interpretation of H$_2$ emissions in extragalactic objects,
because signal to noise limitations, the presence of
bright and uneven stellar continua, and -- particularly in earlier work
-- low spectral resolution, have meant that often only the few
strongest lines could be detected.

Recent results from \cite{gil00} and \cite{dav00} show that these
difficulties can 
in fact be overcome, and that detailed studies of the molecular hydrogen in
extragalactic objects is possible as long as there is sufficient
signal to noise to detect some of the weaker lines.
\cite{gil00} observed individual star clusters in the Antennae
(NGC\,4038/9), which lies at a distance of only 19\,Mpc.
Near the embedded star-cluster located in the mid-IR peak
they detected 22 H$_2$ transitions, 
including some from very high vibrational levels, 
extended over an area twice the
size of the continuum and nebular regions. Gilbert et al.~concluded
that the emission lines are produced by
pure radiative FUV fluorescence,
most likely in PDRs illuminated by the young star clusters.
In work on the two nuclei in Mkn\,266, which at 115\,Mpc
is too distant to permit studies of individual star clusters, \cite{dav00}
used stellar templates to subtract the  continuum,
allowing the fluxes of the weaker lines to be measured.
They found that in neither case could single shock or PDR models fit
the H$_2$ line ratios. They
concluded that $\sim70$\% of the prominent
2.12 $\mu$m 1-0\,S(1) line originates
in non-dissociative shocks; 
and, based on morphology and energetics, that the remaining line
luminosity arises in X-ray irradiated gas, 
FUV-pumped gas, and in dissociative shocks.

In this paper we study H$_2$ excitation in ultraluminous infrared galaxies
(ULIRGs). Such objects often
have 1-0\,S(1) to Br$\gamma$ line
ratios greater than unity \citep{gol95,mur01}.
In star forming regions where the 1-0 emission is
produced by FUV-pumped fluorescence the
1-0\,S(1)/Br$\gamma$ intensity ratio is $\lesssim 1$
\citep[e.g.][]{pux88,doy94,dav98}, suggesting that in ULIRGs 
the 1-0 emission is collisionally excited in hot gas leading
to higher H$_2$ line surface brightnesses.
Indeed, because the majority of ULIRGs show signs of 
elevated star formation rates, as well as interactions and
superwinds \citep{san96}, one might expect shocks or dense PDRs
to play important
roles in the excitation and emission processes.
We have obtained spectra deep enough to detect
$v=2$--1 and 3--2 transitions, with the goal of
constraining the H$_2$
excitation mechanisms in the ULIRGs.

\section{Observations and Data Reduction}
\label{sec:obs}

Our data were obtained as part of the service observing programme at
the {\em Very Large Telescope} between June and September 2001.
ISAAC was used in Medium Resolution mode in the K-band, providing a
nominal resolution of R=2600 with a 1\arcsec\ slit.
Since the wavelength coverage in this mode is only 0.122\,\micron, two
settings were used with an overlap of 5--10\%.
The total integration time was 20\,minutes per setting.
Standard calibrations were performed, including atmospheric standard
stars (type B or G0V), arcs, flatfields, and dark frames.
The data were reduced using PC-IRAF~2.11.3 using standard techniques.
Wavelength calibration was achieved in two steps:
the arc frames were used to apply a transformation which
removed curvature and provided an initial calibration;
then the sky lines on the object frames themselves were used to remove
any residual tilt and fix the final calibration.

Spectra were extracted over 5 pixels (0.75\arcsec).
Although in a number of cases the line emission was extended (in two
cases very much so) beyond this, a short aperture was used for
two reasons:
(1) the slit position angles were assigned rather arbitrarily, and so to
avoid biassing the analysis, we restricted the spectra to the
nuclear region; and
(2) the signal-to-noise was invariably reduced if the spectra were
extracted over a longer aperture, most likely as a result of the
bright continuum.
The two segments of the spectrum for each object were combined after
applying a scaling measured from the overlapping region.
The same scaling was then also applied to the reduced 2D spectra.

An approximate flux calibration was derived from the
standard stars,
the K-band magnitudes of which were estimated from the spectral type
and V-band magnitude;
there was a very good correspondance (to better than 0.1\,mag) to the
K-band magnitudes that could be measured directly (e.g. from
the 2MASS catalogue).
The K-band magnitudes are typically up to
$\sim$1\,mag fainter than values published elsewhere. 
This is a result of the smaller aperture used here
(e.g.~\citealt{gol97} used 3$\times$3 or 3$\times$9 arcsec apertures).
Similarly the 1-0\,S(1) fluxes are typically a factor $\sim 2$ fainter than
those of \citet{gol97}.
Notably, the flux is 3 times fainter in IRAS\,16164$-$0746 and
MCG\,$-$03$-$04$-$014, but position-velocity diagrams for these
objects show the line emission is very extended; 
on the other hand, the flux is only 50\% less for IRAS\,14378$-$3651.

\begin{figure}
\centerline{\psfig{file=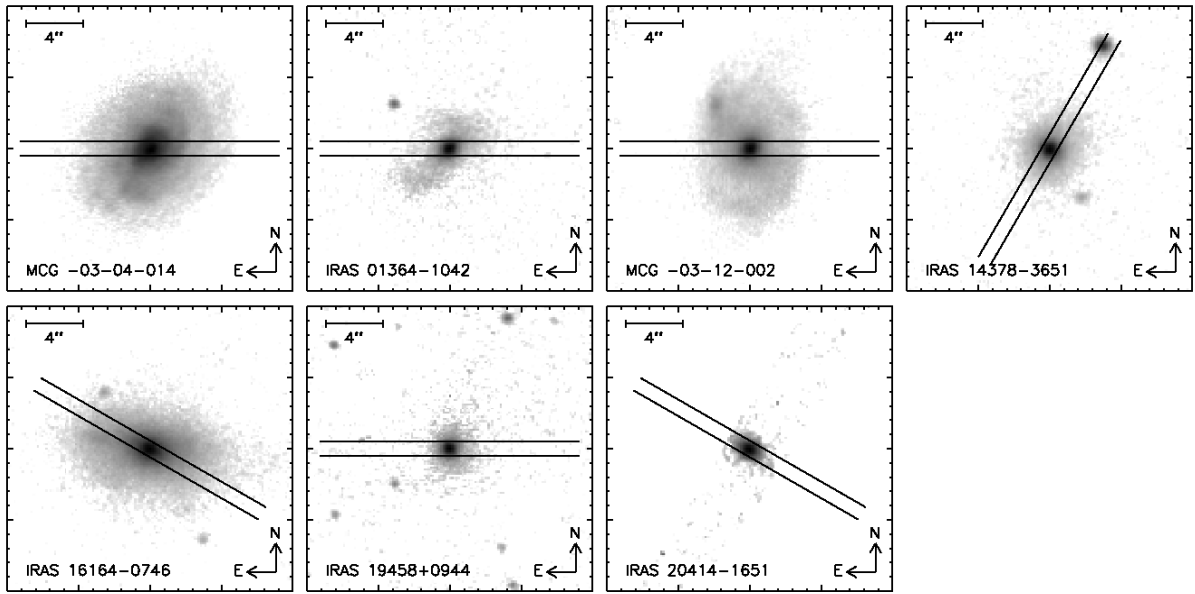,width=16cm}}
\caption{Acquisition images of the targets taken through a $K_s$
filter (with the exception of IRAS\,01364$-$1042 and
IRAS\,20414$-$1651, which were taken through a narrow-band
2.19\,\micron\ filter due to the presence of a bright star in the
ISAAC field of view). 
Each field is 20\arcsec$\times$20\arcsec, and is drawn with
logarithmic scaling.
The parallel lines indicate the position angle and width of the slit.
Due to cosmetic effects, only a narrow strip of the image for
IRAS\,20414$-$1651 is shown.}
\label{fig:acq}
\end{figure}

Acquisition images with integration times of a few seconds were taken
before each spectrum. 
These are shown in Fig.~\ref{fig:acq} together with the orientation of
the slit across the nucleus of each object.
The position angle of the slit was set to lie along the major axis of
the object as determined from 2MASS K-band or red
Digitised Sky Survey images. 
Where there was no clear axis, the slit was set east-west.

\section{Continuum Fitting and Line Extraction}
\label{sec:cont}

\begin{figure}
\centerline{\psfig{file=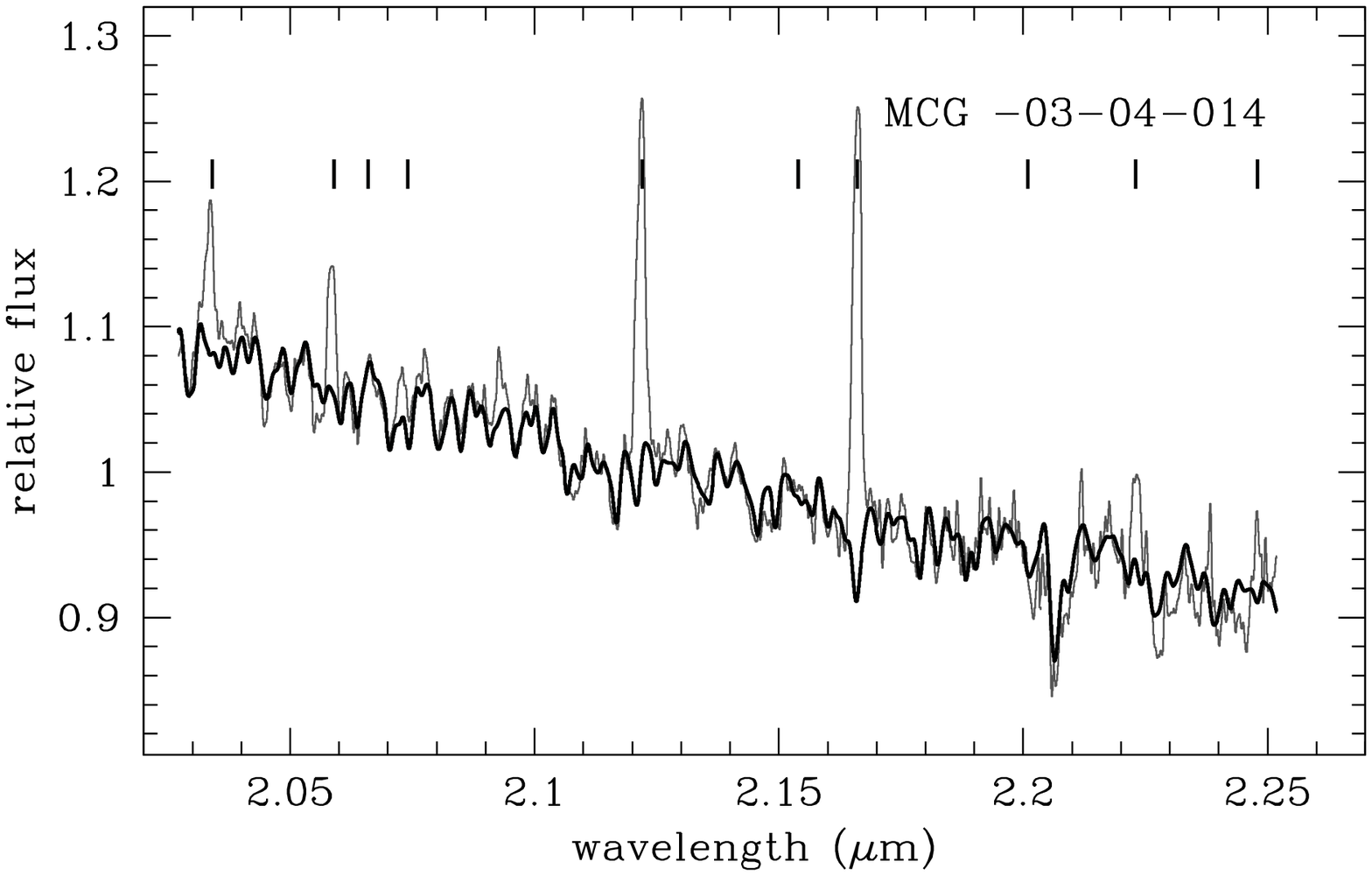,width=8cm}\hspace{5mm}\psfig{file=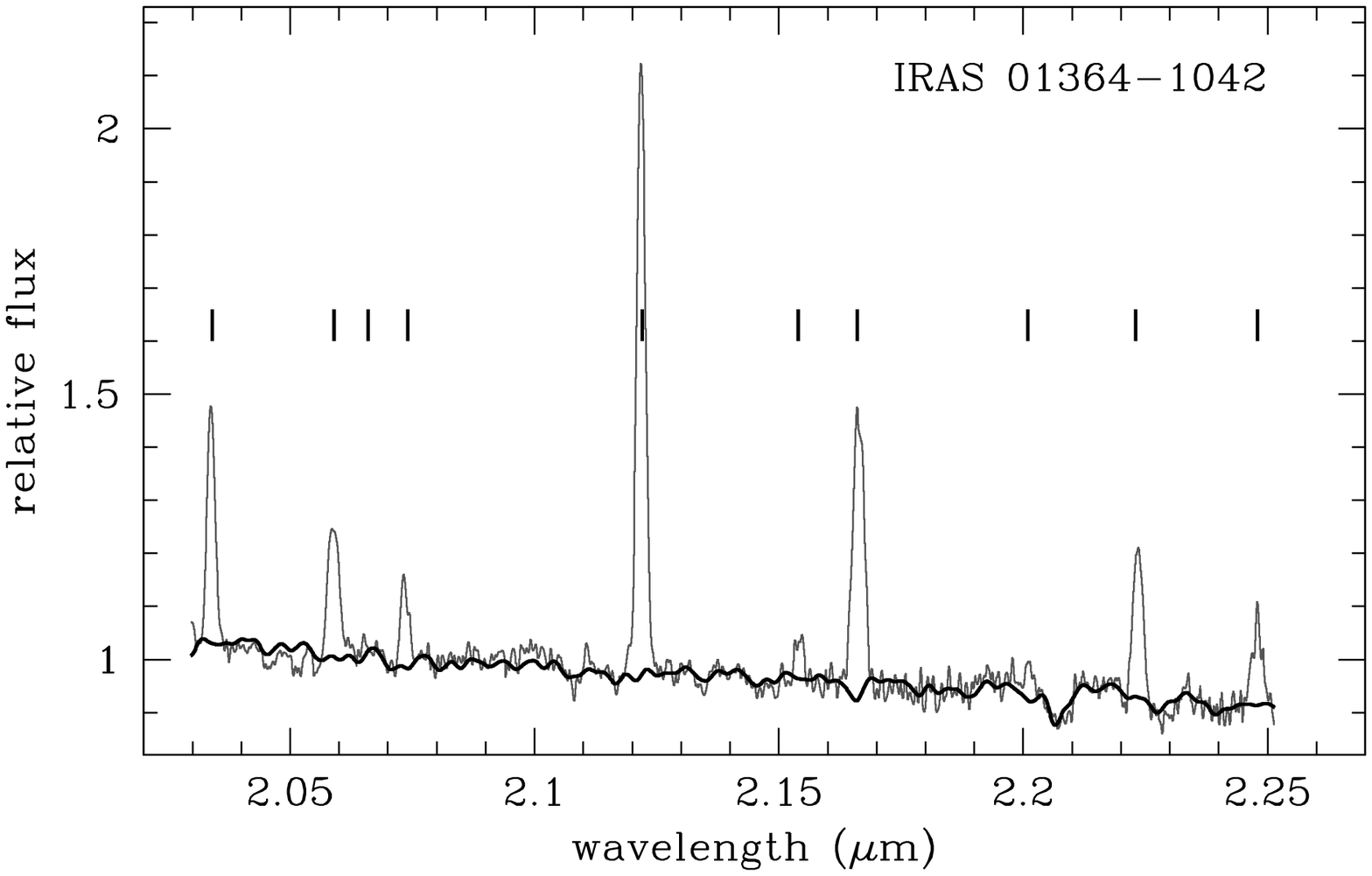,width=8cm}}
\vspace{5mm}
\centerline{\psfig{file=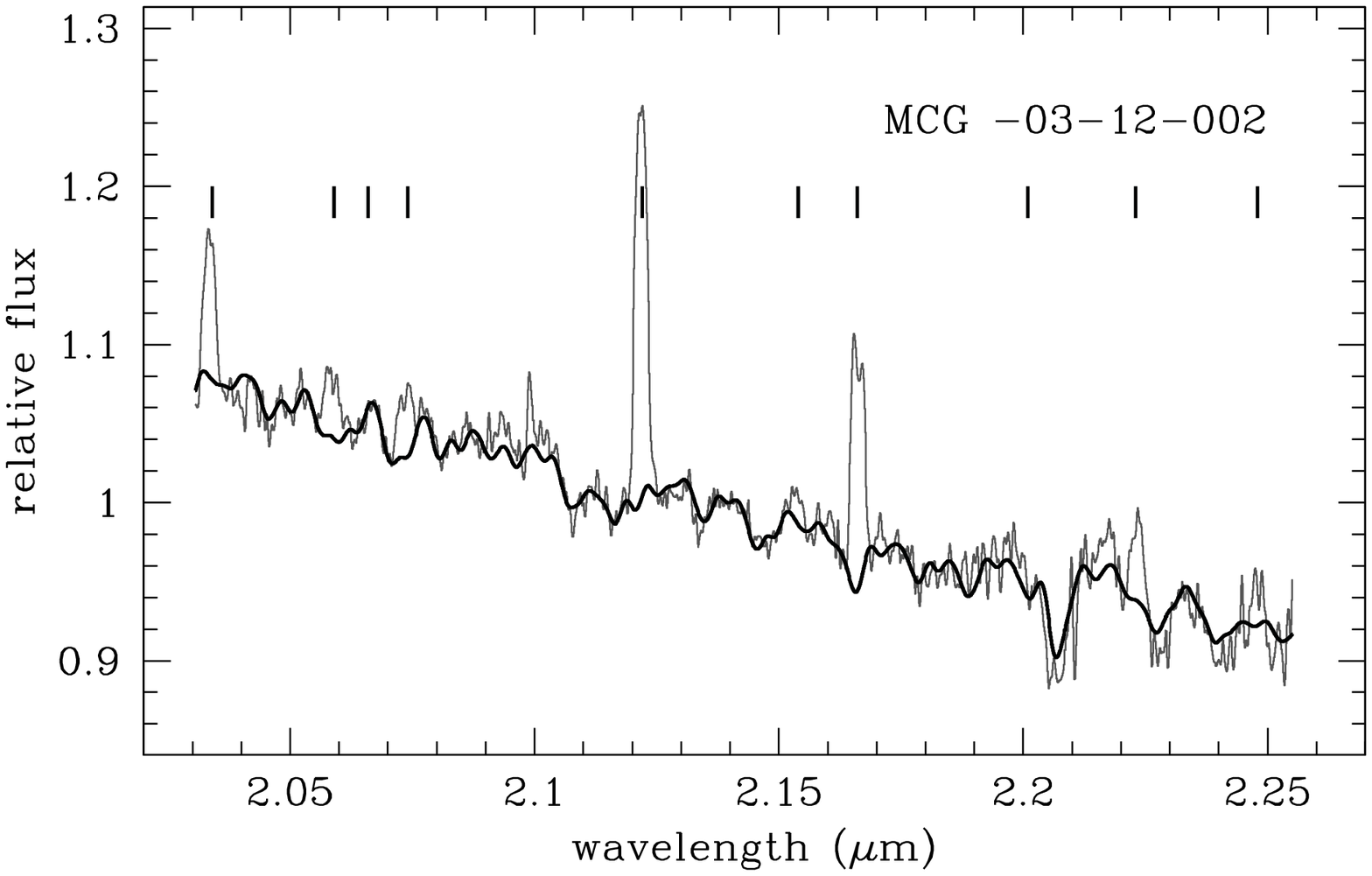,width=8cm}\hspace{5mm}\psfig{file=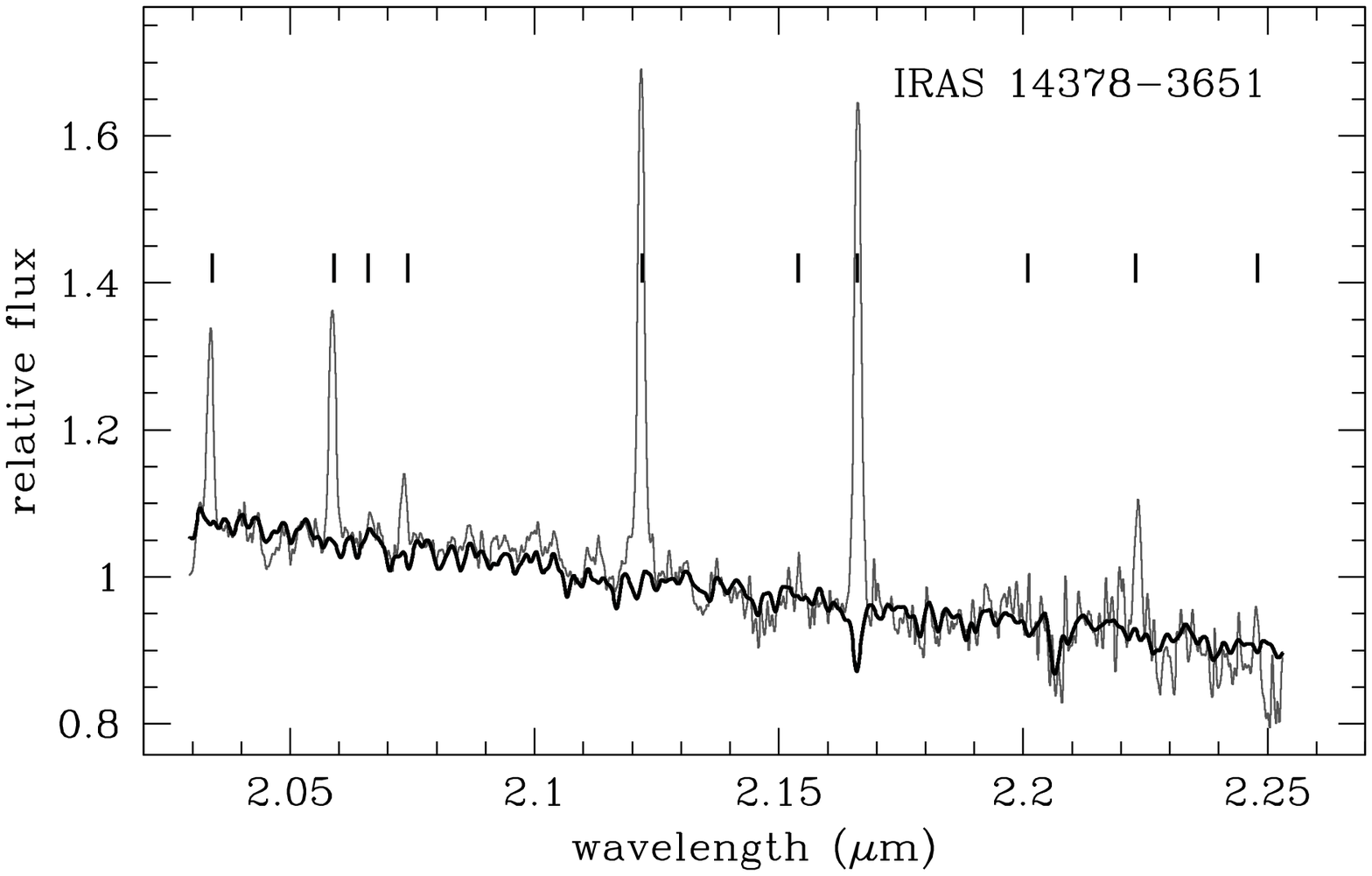,width=8cm}}
\caption{Spectra of each object, plotted at rest wavelength and
normalised to their mean values.
The thick overlaid line is the best fitting continuum constructed from
stellar templates (see text for details).
The vertical marks indicate the positions of the emission lines
recorded in Table~\ref{tab:ratios}. 
Subtraction of the continuum is necessary in order to detect the faint
lines such as 3-2\,S(3) at 2.2014\,\micron, which can be seen clearly in
IRAS\,01364$-$1042.
}
\label{fig:spec}
\end{figure}

The weakest $v$ = 3-2 lines that we have detected 
have strengths that are only a few percent of 
the 1-0\,S(1) line. These are difficult to detect because
they are similar in scale to the absorption features present in the stellar
continua dominated by late type giant or supergiant stars.
In particular the 3-2\,S(5) line at 2.066\,\micron\ coincides with a
bump on the continuum at 2.067\,\micron; 
and the 3-2\,S(3) line at 2.201\,\micron\ sits on the edge of the Na
absorption at 2.207\,\micron.
To overcome these difficulties we have fitted and subtracted a continuum
constructed from the stellar templates of \citet{wal97}.
This library includes over 100 K-band spectra of stars spanning
spectral types O--M and luminosity classes I--V, at a resolution of
R$\sim$3000.
We tried a variety of combinations of template stars, and concluded
that while there was very little difference between the fits found
using giants or supergiants, the dwarfs produced a noticeably poorer
match to the continuum.
Finally we used 7 templates of supergiants ranging in spectral type
from F5 to M5.
Extending the range to earlier types or increasing the number of
templates used had no noticeable effect on the final result.
During the fitting process, the relative scalings of these were
varied to minimise the difference 
between their sum and the spectrum, excluding any regions close to
line emission.
The resulting combination of templates was then reddened (i.e. effectively
allowing adjustments to the overall slope of the continuum) using the
\citet{how83} extinction curve, and
finally convolved with a Gaussian to take account of the stellar
velocity dispersion.
The extinction and Gaussian FWHM were allowed to vary during the
minimisation.
In all cases except for IRAS\,16164$-$0746, the extinction derived in
this way was small ($A_V<1$), a result that is not surprising if the
stars and dust are mixed.
The results of the fitting procedure are shown in Fig.~\ref{fig:spec}
as the thick lines drawn over the spectra:
the fits are successful
and in most cases match the observed continuum features well.
This confirms that the scaling between the two segments of each
spectrum is correct, and that the weak line fluxes are likely to be
accurate.

Line fluxes were measured from the spectra after subtraction of the
stellar continuum, and are given in Table~\ref{tab:ratios}.
We have detected the S(2), S(1)
and S(0) lines in the $v=1$-0 band, the S(3), S(2), and S(1) lines
in the 2-1 band, and 
the S(3) line in the 3-2 band. The transition wavelengths are
listed in Table~\ref{tab:ratios}. We have also set upper
limits for the 3-2 S(5) line
intensities. In addition, in each galaxy
we detected the H{\small \,I} 2.17 $\mu$m
Br$\gamma$ and the He{\small \,I} $2^1P - 2^1S$ 2.06 $\mu$m recombination
lines.  
The uncertainties quoted are measured directly from the spectra as the
{\small rms} of the residual and hence are rather higher than the errors
that would be estimated purely from photon noise, including as they do
the effects of improper line and continuum fitting.
Based on this, we have reached a mean signal-to-noise of 
$25\sigma$ for the 1-0\,S(1) line.

It is apparent from Table~\ref{tab:ratios} that although there is
considerable variation in the ratios of He{\small \,I} and Br$\gamma$ to
1-0\,S(1), the H$_2$ line ratios are in all cases very similar.
Given the range of values which the ratios could take amongst these
objects, the standard deviation in the ratio of any particular H$_2$
line to 1-0\,S(1) of only 0.02--0.03 is small.
We have therefore derived a mean set of ratios, which we also
include in Table~\ref{tab:ratios}.

\section{H$_2$ Excitation Diagrams and PDR Models}
\label{sec:rat}

%\subsection{Excitation Diagrams}
%\label{sec:models}

Given the H$_2$ line fluxes $f$ (W m$^{-2}$) listed
in Table~\ref{tab:ratios}, we have calculated the implied molecular
column densities, $N_{vj}\equiv 4\pi f/A\Omega$, in
the upper rotational-vibrational ($vj$) levels of
the observed transitions, where $A$ is the radiative rate
and $\Omega=0.75\arcsec^2$ is our aperture size. 
We assume that the quadrupole transitions are optically thin
and we use the radiative
$A$ values as given by \citet{wol98}. In 
Figs.~\ref{fig:popmean} and~\ref{fig:pop} we show plots
of $\log{N_{vj}/g_{vj}}$ (where $g_{vj}$ is the statistical weight)
vs.~the energy (in K) of each observed $vj$ level.
We have normalized the population distributions
relative to $N_{1,3}/g_{1,3}$ as inferred from the 1-0 S(1) line.

For thermalised populations at fixed gas
temperature the log$N/g$ points should lie on a straight line
in the excitation diagrams. Clearly,
such single component models
do not fit the data. While the $v=1$ levels appear to
be thermalized at $T\sim 1300$ K, inclusion of the
$v=2$ levels raises the best-fitting temperature to
$T\sim 2400$ K. Even at this temperature, however, the $v=3$ levels remain
underpopulated, and the excitation temperature
inferred from the relative populations in the $v=2$ and $v=3$ levels
exceeds 5000 K. At such gas temperatures the 
molecules would be rapidly dissociated, suggesting that a non-thermal
excitation mechanism is responsible for the excitation of the
$v=2$ and $v=3$ levels.

Collisional excitation in shocks could possibly account for the 1-0
emissions.
A line intensity of $10^{-7}$\,W\,m$^{-2}$\,sr$^{-1}$ is equivalent to
a flux of $2\times10^{-18}$\,W\,m$^{-2}$ for our 0.75\arcsec$^2$
aperture, consistent with the fluxes given in
Table~\ref{tab:measdat}.
It is only in models of non-dissociative C-shocks with velocities
$\gtrsim25$\,km\,s$^{-1}$ and preshock densities
$\gtrsim10^4$\,cm$^{-3}$ that the predicted 1-0\,S(1) intensity
reaches such values \citep{bur90a,tim98,wil00}.
In C-shocks at speeds lower than this, and in all realistic J-shocks,
the 1-0\,S(1) intensity is significantly lower.
In C-shocks at or above this speed (and for gas densities
$>100$\,cm$^{-3}$) the post shock temperatures are of order 1000\,K.
Such models would have difficulty in accounting for the relatively
large columns in the $\nu=2$ and $\nu=3$ levels.
Furthermore, for shocked emission the ortho-to-para ratio in vibrationally
excited levels should attain an equilibrium value of 3 \citep{tim98,wil00}.

%____________________
%However, we show that mechanical energy associated with the star
%formation, primarily via supernova remnants, cannot account for the
%observed line strength.
%\cite{dra90} estimated the 1-0\,S(1) emission that could be
%expected from type {\small ii} supernovae, and derived a conversion
%efficiency of mechanical energy to line emission of
%$6\times10^{-4}$.
%Using this, the ratio of the Lyc and Br$\gamma$ photon rates,
%and the ratio of mechanical to ionising
%luminosity (which initially increases as a function of time, and then
%stabilises) predicted by star formation models, one can
%derive the 1-0\,S(1)/Br$\gamma$ ratio expected 
%if only supernovae and stellar winds contribute to the 1-0\,S(1) flux.
%For continuous star formation models the ionising luminosity
%stabilises after a few Myr;
%the mechanical luminosity also reaches a maximum value, although only
%after a few tens of Myr \citep{lei99}.
%Assuming that star formation has reached this stage, which gives a
%maximum value for the ratio, one finds 1-0\,S(1)/Br$\gamma=0.14$.
%Although there could be a considerable uncertainty in the
%mechanical energy to line emission conversion efficiency above, 
%the ratio is still much less than the range of 0.8--1.6 we have
%observed, and suggests that some other process must dominate the
%1-0\,S(1) emission.
%------------------------

In contrast to shocks, the observed H$_2$ excitation could be 
produced in dense PDRs. In such clouds, FUV irradiation
can heat the gas to temperatures $\sim 1000$\,K, leading
to collisionally excited and thermalized 1-0 line emissions,
while direct FUV-pumping maintains non-thermal 
fluorescent populations in higher-lying vibrational levels
\citep{ste89,bur90b,dra96}.
In fact, a high vibrational excitation temperature, such as
the 5000\,K inferred from the relative $v=3$ and $v=2$ populations,
is a basic characteristic of FUV-pumped fluorescent emission
\citep[]{bla87,ste88}. Furthermore, an additional signature of
fluorescent excitation appears to be present in our data set.
For those objects in which the $v=2-1$
S(1), S(2), and S(3) lines were detected, the resulting
$N/g$ values for the $j=3$, 4, and 5 levels do not appear thermalised.
Instead, the $j=4$ para-level is shifted
above the adjacent $j=3$ and $j=5$ ortho-levels.
Thus, the ortho-to-para ratio 
for the $v=2$ molecules appears
to be out of equilibrium and suppressed below the 
ortho-to-para ratio of 3 obtained in local
thermodynamic equilibrium (LTE) in warm gas. 
This behavior is precisely what is expected 
for excitation by FUV-pumping, 
in which the greater optical depths
in the ortho UV absorption lines supresses the
populations in the vibrationally pumped ortho-H$_2$
relative to para-H$_2$ \citep{ste99}.

%The above behavior, i.e., thermalized rotational populations in $v=1$,
%combined with non-thermal distributions in $v=2$ and $v=3$ characteristic
%of FUV-pumping, is evidence that the H$_2$ emissions in
%our ULIRG sample may originate in dense and hot PDRs. In such systems
%gas heating by stellar FUV radiation fields heat the
%outer PDR layers to temperatures $\gtrsim 10^3$ K leading
%to collisional excitation of
%$v=1$ levels, with continued radiative FUV-pumping of the
%higher lying vibrational levels \citep[]{ste89,bur90}.

We have constructed a series of representative PDR models to
determine PDR conditions and parameters
consistent with the observed H$_2$ line ratios.
For this purpose we have used the code described
in \citet{ste89,ste95}, and \citet{ste99}. 
Briefly, our models consist of
static, plane-parallel, semi-infinite clouds that are
exposed to isotropic FUV radiation fields. At each
cloud depth we compute the equilibrium
atomic to molecular hydrogen density ratio,
$n(H)/n(H_2)$, and we solve for the 
steady-state population densities in the rotational
and vibrational H$_2$ levels in the ground electronic
state. In solving for the $vj$ populations
we include the effects of FUV-pumping via the Lyman
and Werner bands, collisional
processes with H$^+$, H, and H$_2$, 
and quadrupole radiative decays. 

We present a series of five representative PDR models
in which we consider a range of illuminating
FUV field intensities $\chi$,
relative to the
FUV field in the local interstellar medium 
\citep[$2.1\times10^{11}$\,photons\,s$^{-1}$\,m$^{-2}$,][]{dra78},
and total hydrogen particle densities
$n_{\rm H} = n({\rm H}) + 2n({\rm H_2})$ in the clouds.
In all of the models except model 1 (for which we keep the temperature
constant at $T=100$\,K), we assume that the gas temperature varies
with cloud depth as
\begin{equation}
T = {T_{\rm max} \over 1 + 39\times(2n({\rm H_2})/n_{\rm H})^5 }
\label{eq:T}
\end{equation}
where $T_{\rm max}\sim 10^3$ K is the temperature at the cloud edge.
For such thermal profiles the gas temperature equals $T_{\rm max}$
in the outer atomic zone of the PDR, and then declines significantly
to $\sim 25$ K as the gas becomes molecular.
This behavior is consistent with theoretical
expectations for dense PDRs, and is also empirically based
(e.g.~the Galactic star forming cloud S140,
as analyzed by \citet{tim96} and \citet{ste99}).

We summarize the parameters for our five models in
Table~\ref{tab:models}.
The level populations from the models are given in
Table~\ref{tab:ratios} (below the observed values),
and are plotted in Figure~\ref{fig:popmean}.
Also shown in the figure are the populations derived
from the mean observed line ratios. We now give
a short summary of the models and how well they describe the mean
observed line ratios:
\begin{description}
\item[model 1]
Low density, $n_{\rm H}=10^3$\,cm$^{-3}$, and cool (isothermal) 
$T=100$ K gas, excited by a relatively weak $\chi=10^2$ FUV field. 
This results in low surface brightness H$_2$ emission,
with $I_{\rm 1-0 S(1)}=1.3\times 10^{-9}$ W m$^{-2}$ sr$^{-1}$.
The population ratios, including the $v=1$ levels
are characteristic of pure radiative fluorescent excitation,
unaffected by vibrational collisional excitation or
deexcitation.  This model does not produce a good fit to the data.
\item[model 2]
The gas density and UV field are both increased by an order of
magnitude to $n_{\rm H}=10^4$\,cm$^{-3}$ and $\chi=10^3$, 
and a thermal profile as given by Eq.~\ref{eq:T} is adopted
with $T_{\rm max}=10^3$\,K.
This results in much higher line surface brightnesses,
especially for the 1-0 transitions, with  
$I_{\rm 1-0 S(1)}=5.2\times 10^{-8}$ W m$^{-2}$ sr$^{-1}$.
The $v=1$ levels are
thermalised by collisions, while the $v=2\ \&\ 3$ levels
are excited by FUV-pumping and therefore
display the ortho-to-para
shifts characteristic of fluorescent excitation \citep{ste99}.
This model matches the data rather well, showing that 
the H$_2$ populations are consistent with those expected
for dense PDRs with hot outer boundaries.
\item[model 3]
Similar to model 2, with again $n_{\rm H}=10^4$\,cm$^{-3}$ and $\chi=10^3$,
but with $T_{\rm max}$ increased to 2000\,K.
This increases the surface brightnesses in the 1-0 lines still further,
with $I_{\rm 1-0 S(1)}=2.0\times 10^{-6}$ W m$^{-2}$ sr$^{-1}$.
The higher $v$ levels are now also thermalised and therefore
have much lower
populations relative to the $v=1$ level.
The result is very similar to a simple purely ``thermal'' model
in which the populations are assumed to be in LTE at
$2000$\,K, and very
much under-predicts the $v=2\ \&\ 3$ populations.
\item[model 4]
Also similar to model 2, with $n_{\rm H}=10^4$\,cm$^{-3}$ and 
$T_{\rm max}=10^3$\,K, but with the UV field increased by a factor of 100
to $\chi=10^5$.
The relative populations and line intensities are hardly
affected compared to model 2, and
there is only a slight increase in the populations of the higher $v$ levels.
\item[model 5]
As in model 4, having $\chi=10^5$ and $T_{\rm max}=10^3$\,K, but 
with the gas density increased by a factor of 100 to 
$n_{\rm H}=10^6$\,cm$^{-3}$. The higher density leads to more
effective thermalisation in the higher $v$
levels, so that these levels are under predicted by the model,
although not as severely as in model 3.
\end{description}

It is clear that models 2 and 4 provide the best match to the mean
line ratios, and that with the current data they are effectively
indestinguishable.
The 1-0\,S(1) line intensities of the 2 models are also similar, and
consistent with those observed in the ULIRGs.

In our models we have not explicitly computed
the heating and cooling balance that would be expected to
yield hot gas in the outer photodissociated atomic zones.
In PDR theory the 6--13.6\,eV photons are available for
heating via photoelectric emission from dust grains, and
collisional deexcitation of FUV-pumped H$_2$. 
For our model 2, the total energy flux in this band is
$2.6\times10^{-3}$\,W\,m$^{-2}$;
while the emergent flux in the H$_2$ 1-0 S(1) line is
$5.2\times10^{-8}$\,W\,m$^{-2}$, or
$2\times 10^{-5}$ of the available FUV energy, a fraction
consistent with
theoretical computations \citep{ste89,bur90b}.
Thus it is possible that the H$_2$ emission -- including the
thermalised $v=1-0$ emissions -- is produced entirely in PDRs.
However, for the 1-0 lines we cannot 
rule out contributions from additional sources
such as shock waves.

Having looked at which models might match the mean ratios, we 
now briefly consider
the individual objects.

From surface brightness considerations alone, model 1 can be ruled out
for all objects.
Integrated over the 0.75\arcsec$^2$ aperture in which we have measured
the line fluxes, this model would predict a maximum 1-0\,S(1) flux of
$2.3\times10^{-20}$\,W\,m$^{-2}$, nearly 2 orders of magnitude less
than the range observed in our sample.
We would also argue against model 3 because it requires that
all the levels are effectively thermalised at the same temperature. 
We now consider the remainder of the models for each object.
For IRAS\,19458$+$0944, models 2 \& 4 provide good fits to the data,
and while model 5 also fits it is more to the extreme of the
errorbars.
These three models also match the data for MCG\,$-$03$-$12$-$002 well,
although the $v=1$ levels indicate a temperature at the edge
of the clouds somewhat warmer than 1000\,K.
For MCG\,$-$03$-$04$-$014 models 2 and
4 also match the data well, but here the extra data
point for $v=2$ levels suggests that model 5 is less likely.
IRAS\,01364$-$1042 and IRAS\,16164$-$0746 present some of the best data.
For these objects model 5 van be confidentally 
dismissed in favour of models 2 and 4.
For the former, the $v=1$ populations suggest, as for
MCG\,$-$03$-$12$-$002, that the temperature at the edge of the clouds
is slightly warmer than 1000\,K.
The cases for IRAS\,14378$-$3651 and IRAS\,20414$-$1651 are rather
uncertain and models 2, 4, and 5 all match the data well;
although in the latter object, there is a tendency towards the higher
density model since the $v=2$ distributions appear more thermalised.

%We have considered models of clumpy PDRs, in which 2 or more
%components are present, characterised by different beam filling
%factors \cite[c.f.][]{bur90b}.
%Given the data presented here, a full discussion of this possibility
%is not warranted.
%However, we can hypothesise that towards the inner regions there are
%dense strongly irradiated PDRs, while further away from the nucleus
%itself the gas is less dense and the FUV flux less intense.
%The dense components could populate the $\nu=1-0$ levels, as in 
%model 3;
%but low density PDRs irradiated by weak FUV fields which are able to
%populate the higher vibrational levels would also have a strong effect
%on the $\nu=1-0$ levels, as seen in model 1.
%In the data, the $\nu=1-0$ levels are well thermalised, suggesting
%that there is no significant contribution to the H$_2$ emission from
%low density PDRs.
%However, PDRs with different characteristics are
%possible (and by analogy to Orion perhaps likely, see
%Section~\ref{sec:comp}), as long as the total H$_2$ emission they
%produce contributes only a small fraction of the total.

Our conclusion from the line intensities and
ratios is that all the objects can be characterised by 
simple single-component PDRs
illuminated by FUV fields 
with $\chi \geqslant 10^3$, and hydrogen densities in the range 
n$_{\rm H} = 10^4$--$10^5$\,cm$^{-3}$ (possibly in some cases as
high as $10^6$\,cm$^{-3}$), and a temperature at the outer edge of the
clouds $T=1000$--1500\,K. Multi-component models \citep{bur90b} could also
be considered, in which, for example, an inner dense region
is responsible for the thermalised 1-0 emissions, with contributions
to the fluorescent component from more extended low density gas
(see also our discussion of Orion in \S 4.2). However, given
the limited spatial information in our data set we do
not consider such models here.

Observations of additional PDR 
diagnostic lines such as [O{\small \,II}] 63\,$\mu$m and
145\,$\mu$m, [C{\small \,II}] 158\,$\mu$m, or high-J CO transitions
\citep{tie85,ste89,koe94}
would provide valuable additional constraints on the PDR conditions.
However, no such data is yet available for our sources.

\subsection{Starburst Models}
\label{sec:models}

If the observed molecular hydrogen emissions are produced
in PDRs then it is of interest to consider whether the FUV sources
-- most likely hot OB stars -- can also provide the 
Lyman continuum (Lyc; $h\nu > 13.6$ eV) EUV photons
responsible for the Br$\gamma$ recombination
line emission.
As we will now show, the relative photon production rates
in the FUV and EUV bands that are implied by the H$_2$ and Br$\gamma$ 
line fluxes, are consistent with 
young star clusters containing populations of massive hot stars.

In Table~\ref{tab:Qratio} we present the results of starburst model
computations using our starburst code STARS
\citep[e.g.][]{ste98,tho00} for the Lyc and FUV-band photon rates
$Q_{\rm Lyc}$ and $Q_{\rm H_2}$ produced in evolving star clusters.
We define the FUV-band as spanning
912 to 1130\,\AA, since photons in this wavelength range can be
absorbed in the H$_2$ Lyman and Werner absorption line systems,
leading to FUV-pumped fluorescent H$_2$ emission lines.
In Table~\ref{tab:Qratio} we also list the helium ionizing 
(He{\small \,I}; $h\nu > 24$ eV) photon production rates, and the associated
K-band luminosities, $L_K$\,\footnote{we define $L_K$ as the total
luminosity in the 
1.9--2.5\,\micron\ K-band for a frequency independent K-band flux
density. If $S_{2.2}$ is the 2.2\,\micron\ flux density in mJy and
$D_{\rm Mpc}$ is the source distance in Mpc then 
$L_K (L_\odot) = 1.17\times10^4 D^2_{\rm Mpc} S_{2.2}$. The
absolute K-magnitude $M_K = -0.33 -2.5\log{L_K}$},
and bolometric luminosities
$L_{\rm bol}$ of the model clusters.  We consider a
range of cluster ages, for both instantaneous ($10^5$ $M_\odot$) bursts and 
continuous ($SFR=1$ $M_\odot$ yr$^{-1}$) star formation,  
assuming Salpeter initial mass functions
(IMFs) with upper-mass limits of $M_{\rm up}=30$ and 120 $M_\odot$.
Our starburst models incorporate hot-star
``wind-atmospheres'' in the computation of the cluster luminosities and
spectral energy distributions \citep{ste03}.
We assume solar metallicity for the evolutionary tracks and
stellar atmospheres.  

Table~\ref{tab:Qratio} shows that the ratio $Q_{\rm H_2}/Q_{\rm Lyc}$ varies
with cluster age and also depends on $M_{\rm up}$. For instantaneous
bursts, $Q_{\rm H_2}/Q_{\rm Lyc}$ increases rapidly with age
as the hottest Lyc producing O stars disappear from the system,
whereas FUV photons continue to be produced by the B-stars.
For continuous star formation $Q_{\rm H_2}/Q_{\rm Lyc}$  increases
slightly with time and reaches fixed values as the
relative numbers of O and B type stars reach equilibrium.
At early times the photon ratio is smaller for larger $M_{\rm up}$
due to the efficient production of Lyc photons in the most
massive stars.

We determine $Q_{\rm Lyc}$ for our sources from the
Br$\gamma$ fluxes assuming case B recombination at $10^4$\,K,
for which 71.9 Br$\gamma$
photons are produced for every Lyc photon 
absorbed in photoionizations. Similarly, $Q_{\rm H_2}$ is
proportional to the flux in a FUV-pumped fluorescent H$_2$
emission line. To estimate $Q_{\rm H_2}$ we use the flux in the  
the 2-1\,S(3) line since this is a FUV-pumped line that
we detected in every galaxy in our sample. The FUV-band to H$_2$
line photon ``conversion efficiency'' depends on the
ratio of the incident FUV intensity to gas density, $\chi/n_{\rm H}$,
a PDR parameter that controls the fraction of the incident
FUV photons absorbed by molecules as opposed to
dust grains in the PDRs \citep{bla87,ste88}.
The efficiency is largest when $\chi/n_{\rm H} \lesssim 0.01$ cm$^{3}$,
and decreases for larger values of the FUV to density ratio
\footnote{The efficiency also depends on the effective FUV grain
absorption cross section in the PDRs. We assume a cross section
equal to $1.9\times 10^{-21}$ cm$^{-2}$ as appropriate
for Galactic dust \citep{dra03}.}. % astro-ph/0304489
We assume a FUV to 2-1\,S(3) conversion factor of
$\eta = 4.1\times10^{-3}$, as given by our PDR model 2.
For this model $\eta$ is at about half the maximum efficiency.
It follows that the distance independent ratio
\begin{equation}
\frac{Q_{\rm H_2}}{Q_{\rm Lyc}} \ = \ 
	\frac{F_{\rm 2-1\,S(3)}}{F_{\rm Br\gamma}} \ 
	\frac{1}{71.9 \ \eta} \ 
	\frac{\lambda_{2-1\,S(3)}}{\lambda_{Br\gamma}}
	= \ 3.25 \ 
	\frac{F_{\rm 2-1\,S(3)}}{F_{\rm Br\gamma}}.
\end{equation}
In Table~\ref{tab:derdat} we list the FUV to Lyc photon ratios,
that we infer for our ULIRG sample. The ratios range
from 0.3 to 0.9, with 
$Q_{\rm H_2}/Q_{\rm Lyc} \approx 0.5$ a representative value.

Our starburst models thus show that the H$_2$
and Br$\gamma$ line emissions may be produced by
the same populations of
star clusters in the ULIRG nuclei. If the
hydrogen molecules in the PDRs are absorbing 
the FUV photons at close to maximum efficiency,
then the clusters must be young, as shown by the photon ratios listed
in Table~\ref{tab:Qratio}. The clusters must then also
contain massive stars with $M_{\rm up}$ close to 120\,$M_\odot$.
For instantaneous bursts, a ratio $Q_{\rm H_2}/Q_{\rm Lyc} \approx 0.5$
implies an age of $\sim 1$\,Myr. For continuous star formation
the limit on the age is less severe, and could be as large
as $\sim 10$\,Myr, especially if the FUV to H$_2$ line
conversion efficiency is smaller than we have assumed.
It appears then that the same stellar radiation 
sources can plausibly account
for both the H$_2$ lines and the recombination line emissions. This
would not be the case if, for example, the Br$\gamma$/H$_2$
line intensity ratios were much larger than observed.

The 2.06\,$\mu$m He{\small \,I} recombination line provides an additional
constraint, as this line is proportional to $Q_{\rm HeI}$, the
number of He{\small \,I} continuum photons ($h\nu > 24.6$\,eV) produced
per unit time. 
The strong 2.06\,$\mu$m line in our spectra is a further indication of the
presence of very hot and massive stars. However, as noted by \citet{shi93}
converting the helium line flux to $Q_{\rm HeI}$ depends on 
a variety of largely uncertain parameters for the H{\small \,II} regions,
including the nebular densities, gas filling factors, 
and dust content. Here we derive a lower limit, 
$Q^{\rm min}_{\rm He}$, assuming that all of the He{\small \,I} continuum
photons are absorbed in helium ionizations, followed
by (case-B) recombinations. The effective
recombination coefficient for the 2.06\,$\mu$m line 
\citep{smi91,ost89} ranges from
$\alpha^{\rm eff}_{2.06}=4.74\times10^{-14}$\,cm$^3$\,s$^{-1}$
at low densities ($10^{2}$\,cm$^{-3}$), to 
$\alpha^{\rm eff}_{2.06}=6.78\times10^{-14}$\,cm$^3$\,s$^{-1}$
at high densities ($10^{4}$\,cm$^{-3}$), where collisional
depopulation of the $2^3S$ level to the $2^1S$ and $2^1P$ levels
(rather than photon decay to $1^1S$)
enhances the He{\small \,I} 2.06\,\micron\ intensity.
Given a total case-B recombination  
$\alpha_B{\rm (HeI)} = 2.73\times10^{-13}$\,cm$^3$\,s$^{-1}$
\citep{ost89}, it follows that at least 5.8 2.06\,$\mu$m line photons
are emitted per He{\small \,I} ionizing photon
\footnote{We note that since each He recombination results
in about one photon that can ionize H, we have assumed that
our estimates of $Q_{\rm Lyc}$ are unaffected by the presence
of He.}. The resulting limits, $Q^{\rm min}_{\rm He}/Q_{\rm Lyc}$, are
listed in Table~\ref{tab:derdat}. They are consistent with
the young ages we have inferred for the starbursts.

Without any extinction corrections,
the Br$\gamma$ fluxes and source distances imply
Lyc emission rates ranging from $6\times 10^{52}$\,s$^{-1}$ (in
MCG-03-12-002), to $3.4\times 10^{54}$\,s$^{-1}$ (in
IRAS\,19458$+$0944). For young 1 Myr instantaneous bursts this implies
total stellar masses 
ranging from $6.7\times 10^5$ to $3.8\times 10^7$\,M$_\odot$
for a Salpeter IMF ranging from 0.1 to 120\,M$_\odot$.
The implied stellar masses increase significantly if
such clusters are more evolved \cite[c.f.][]{tho00}.
For continuous star formation, the Br$\gamma$ fluxes imply
star formation rates from 0.7 to 38\,M$_\odot$\,yr$^{-1}$,
assuming an age of 5\,Myr. It follows from the starburst
data in Table~\ref{tab:Qratio} that for young (1\,Myr)
instantaneous bursts the predicted K-band luminosities of
clusters account for at most a few percent of the observed
K-band luminosities. For continuous star formation, the
fractions could be a bit larger, up to $\sim20\%$, but are still small.
We conclude that the observed K-band continua are not produced in the
current star forming episode.

Equivalently one can consider the Br$\gamma$ equivalent widths, which
should exceed 100\,\AA\ or more for star formation with ages of
10\,Myr or less.
The directly measured equivalent widths, however, lie in the range
5--20\,\AA\ for these ULIRGs.
Low Br$\gamma$ equivalent widths have also been observed by other
authors \citep{gol95,mur01}.
The depth of the continuum absorption features (in
Section~\ref{sec:cont} we showed that the scale of these features was
similar to that in late-type giants and supergiants)
suggest that it is unlikely that the K-band continua
arise from hot dust associated either with starbursts or hidden
AGN.
On the other hand, it is plausible that the continua are produced
by an older underlying population.
The only firm conclusion we can draw about such a population
is that it contributes little to the total FUV and ionising flux,
which means that either it is at a low level or has ceased in the
recent past.
For evolved 10$^{10}$\,yr stellar populations, M/L$_K\approx20$
(\citealt{thr88}; Fig~3 of \citealt{tac96}), 
implying stellar masses of order 10$^{10}$\,M$_\odot$.
If the near-infrared continuum is dominated by a younger 10$^8$\,yr
population, consistent with that
over which star formation occurs in the major merger models of
\citet{mih96}, then the mass to light ratio is M/L$_K\approx0.7$
\citep{tac96}, implying
much more moderate masses in the range 10$^8$--10$^9$\,M$_\odot$.
In this case, the star formation must have ceased already, since it
would otherwise make a significant contribution to the observed FUV
and ionising fluxes.
We note that in some models of \citet{mih96} (notably the bulgeless
mergers), multiple separate epochs of star formation can be triggered
during an interaction, occuring at each point of closest approach as
the two progenitor galaxies orbit each other before finally merging.

From Tables~\ref{tab:derdat} and~\ref{tab:Qratio} it can also be seen
that the bolometric luminosities based on the star formation rates
implied by the Br$\gamma$ fluxes are very much less than those
inferred by the IRAS fluxes.
The total correction factor needed to reconcile them is in the range 20-40.
The difference could arise from a combination of aperture effects
(discussed in Section~\ref{sec:obs}), extinction, and perhaps also
dust absorption within the nebulae as proposed for ULIRGs by
\cite{voi92} and \cite{bot98}.
Mid infrared spectroscopy by \cite{gen98} indicates that
values of $A_V=5$--50 (screen model) or $A_V=50$--1000 (mixed model)
are not uncommon -- i.e. for lines in the near infrared one may be
seeing as little as 1--10\% of the total flux.
Additionally, dust within the ionised nebulae could absorb the Lyc photons.
FUV photons would also be absorbed,
reducing the FUV intensity at the PDR boundaries.
Because the dust absorption and aperture effects can apply to the line
emitting regions and older stellar population differently, it is not
straight forward to correct for them and we have not attempted to do
so.

Assuming that the observed H$_2$ emission is distributed uniformly
over the 0.75\arcsec$^2$ aperture, the area averaged FUV intensities
within this region are given by
\begin{equation}
\langle \chi \rangle = 
1700 
\biggl({Q_{H_2} \over 10^{52} \ {\rm s^{-1}}}\biggr)
\biggl({R\over 100 \ {\rm pc}}\biggr)^{-2}
\end{equation}
where $R$ is the radius probed by our observational aperture.
For our ULIRG sample, $\langle\chi\rangle$ ranges from 
700 to 2000, broadly consistent with
our PDR model 2.

The FUV fluxes 
could be substantially larger if the filling factor
of the ionized and PDR gas is small. For example, we consider
a collection of identical and young star clusters each
surrounded by its own H{\small \,II} region (which does not overlap with any
other), with PDRs at the outer boundaries of these H{\small \,II}
regions. For such a system, the total Lyc emission rate 
$Q_{\rm Lyc} =  N_{\rm clus} (4\pi/3) \alpha_B n_e^2 r_s^3$
where $N_{\rm clus}$ is the number of star clusters, $n_e$ is the
electron density in the H{\small \,II} regions, $r_s$ the
Str\"omgren radius, and $\alpha_B$ is the hydrogen 
recombination coefficient. The FUV-band flux 
at the edges of the PDRs is $(Q_{\rm H_2}/N_{\rm clus})/4\pi r_s^2$,
so that the FUV flux (in units of the \citet{dra78} field) may be written as
\begin{equation}
\chi = 3.5\times 10^3
\left( \frac{Q_{\rm H_2}}{Q_{\rm Lyc}} \right) \ 
\left( \frac{Q_{\rm Lyc}/N_{\rm clus}}{10^{52}\,{\rm s^{-1}}} \right)^{1/3} \ 
\left( \frac{n_e}{100\,{\rm cm^{-3}}} \right)^{4/3} \ \ \ .
\end{equation}

To estimate the mass of a typical star cluster, we consider the sample of 9
starburst galaxies studied by \citet{meu95} (see also \citealt{mao01}).
They found that the compact star clusters had minimum masses in the
range $6\times10^3$--$2.4\times10^5$\,M$_\odot$ assuming a Salpeter IMF in
the range 5--100\,M$_\odot$.
By extending the lower mass limit to 0.1\,M$_\odot$, this range
increases to $3.3\times10^4$--$1.3\times10^6$\,M$_\odot$.
We adopt $10^5$\,M$_\odot$ as representative of the typical
mass.
For clusters of this mass the Lyc production rate per
cluster at 1\,Myr is $Q_{\rm Lyc}/N_{\rm clus}=9\times 10^{51}$ s$^{-1}$.
For $Q_{\rm H_2}/Q_{\rm Lyc}=0.5$, and $n_e$ between 100 and 10$^3$ cm$^{-3}$
the FUV flux ranges from $2\times 10^3$ to $4\times 10^4$.

We note that the electron densities in ULIRGs reported by \cite{vei99} have a
distribution which peaks at values around 100--200\,cm$^{-3}$ with a
long tail to higher values, and a few above 1000\,cm$^{-3}$.
However, it is not clear whether these densities, which are derived
from the [S{\small \,II}]$\lambda6716$/$\lambda6731$ doublet measured in a
$\sim2\times4$\,kpc aperture, do trace the 
density in the inner star forming regions or whether they are dominated
by more extended emission.
In the latter case, the emission could arise in diffuse gas at
relatively large radii, e.g.~in superwinds, 
and would thus provide only a lower limit on the central density.
The density profiles for a number of galaxies with superwinds have
been measured by \cite{hec90}.
They find that at large radii, greater than a few kpc for the more
luminous galaxies (which include several ULIRGs), the density is
consistent with a $r^{-2}$ 
profile and has a value of 100\,cm$^{-3}$ at some point typically
within $r\sim1$--3\,kpc. 
But $n_e$ attains a more constant value in the range
500--1000\,cm$^{-3}$ at smaller radii. 
%The turnover radius appears to be related to the size of the star
%forming region which they estimate from the momentum
%injection rate and central pressure to range from 300\,pc for M\,82 to
%$\sim2$\,kpc for the most luminous galaxies;
%values which agree well with the observed sizes.
%We conclude that it is only the higher electron densities measured by
%\cite{vei99} which actually trace the properties in the
%central star forming regions of ULIRGs.

In summary we conclude that our Br$\gamma$ and H$_2$ line fluxes
are consistent with nuclear star forming clusters containing
massive stars, with ages $\sim 1$\,Myr for instantaneous bursts,
or more evolved 5--10\,Myr systems for continuous star formation.
The H$_2$ line fluxes, and corresponding FUV-band photon production
rates, are consistent with the intense FUV fluxes, 
$\chi=10^3$--10$^5$ we adopted in our PDR models 2, 4, and 5.
As we have seen, the H$_2$ line ratios imply that the PDRs are dense,
with $n_{\rm H}$ between 10$^4$ and 10$^6$\,cm$^{-3}$,
and hot, with $T=1000$--1500\,K in the outer edges.

\subsection{Comparison of PDRs to Other Star Forming Regions}
\label{sec:comp}

How do the physical conditions for the PDRs -- UV intensity,
temperature, density -- that we have derived in
Section~\ref{sec:models} compare to those in
other star forming regions?

One of the best studied PDRs is that in Orion.
Several authors \citep[e.g.][]{luh98,mar98,usu96} have found that in
some regions of Orion there is clear evidence for dense PDRs
illuminated by intense UV radiation.
The integrated 1-0\,S(1) luminosity within a few arcmin of
$\theta^1$~Ori~C, which includes all these dense PDRs as well
as some outflows with shocked emission, is only 4.9\,L$_\odot$
\citep{usu96}.
However, large scale observations of the Orion cloud suggest that the
total 1-0\,S(1) luminosity is at least 34\,L$_\odot$ and over the
entire $\sim3^\circ\times10^\circ$ CO cloud may be as much as
63\,L$_\odot$ \citep{luh94}.
That is, the extended emission dominates the total emission by a
factor of 7--13.
Based on the surface brightness
($1.4\times10^{-9}$\,W\,m$^{-2}$\,sr$^{-1}$) and the strength of the
6-4\,Q(1) line at 1.601\,\micron\ 
(1-0\,S(1)/6-4\,Q(1)$=2.9$) in the extended emission, the authors
concluded that the excitation mechanism is fluorescence by a weak UV
field in low density PDRs, similar to our PDR model 1.
Hence, although Orion does contain a number of dense intensely
illuminated PDRs, on a global scale -- i.e. if it  were observed from
far enough away to be spatially unresolved -- its measured properties
would be those of the extended emission.
Hence, it does {\em not} appear to be similar to the PDRs in ULIRGs.
By analogy, one would might speculate that in ULIRGs there are some
PDRs which have much greater (or smaller) densities and incident UV
fields than 
the global properties we have been able to measure, but
that these account for only a small fraction of the
total H$_2$ luminosity.

The extended emission in the inner 400\,pc of the Galaxy is different
to that in Orion.
\cite{pak96} found that although there is very intense emission close
to Sgr\,A, the emission extended on degree scales has a lower surface
brightness of $\sim3\times10^{-8}$\,W\,m$^{-2}$\,sr$^{-1}$ (corrected
for foreground extinction of $A_K=2.5$), although
this is still an order of magnitude higher than that in Orion.
Based on this and the uniformity of the emission, the authors
concluded that it was excited in moderately dense PDRs with moderate
UV field -- similar to our PDR model 2, which has a similar density
to but a lower UV intensity than we have observed in ULIRGs.
Table~\ref{tab:models} shows that the intrinsic surface brightness of
1-0\,S(1) emission is similar in both cases.

The 400\,pc scale over which these measurements were made is similar
to that covered by our 0.75\arcsec\ aperture at the distance of the
ULIRGs.
The 1-0\,S(1) line luminosities for the ULIRGs given
in Table~\ref{tab:measdat} has a mean of
$2\times10^6$\,L$_\odot$ without extinction or aperture corrections.
Including this correction factor, we find that the line luminosity is
3--4 orders of magnitude 
greater than the 8000\,L$_\odot$ (already corrected for extinction)
measured in the Galactic Center \citep{pak96}.
Since we have shown that the surface brightness of the individual PDRs
is similar both in the Galactic Center and in ULIRGs, the difference
in total H$_2$ luminosity must be due to a vast increase in the
filling factor of the PDRs.

We argue that this is due not only to an increase in the number of
molecular clouds, but also an increase in the number of molecular
clouds that have become PDRs.
The gas mass in the inner 500--600\,pc of the Galaxy has been determined
by a variety of means including $^{12}$CO(1-0) and C$^{18}$O(1-0) radio
measurements, thermal dust emission detected by IRAS and COBE, and
0.1--1.0\,GeV $\gamma$-rays \cite[see][and references therein]{dah98}.
These all yield similar values, for which the authors estimated the
weighted mean to be $3\times10^7$\,M$_\odot$.
Typically, ULIRGs have a strong central concentration of molecular gas
on the order of $10^{9-10}$\,M$_\odot$, for example as deduced from
observations of the CO(1-0) line by
\cite{eva02} with a 2--4\arcsec\ beam and for a much larger sample,
but also with a larger beam of 13--22\arcsec, by \cite{sol97}.
For the 3 ULIRGs in this paper for which the radio CO\,1-0
luminosities have been measured, the derived gas masses are
1.5--$2.0\times10^{10}$\,M$_\odot$ \citep{mir90}.
These results indicate that the ratio $L_{\rm 1-0\,S(1)}$/$M_{\rm
gas}$ is 10--100 times greater in ULIRGs than in the Galactic Center,
implying a much higher star formation efficiency for the ULIRGs 
\cite[c.f.][and references therein]{san96}.

%We have seen that in both ULIRGs and the inner few hundred pc of the
%Galaxy, the 1-0\,S(1) is predominantly from PDRs.
%And although the UV intensity on the PDRs in ULIRGs is greater, their gas
%densities and 1-0\,S(1) surface brightnesses are similar to those in
%the Galactic Center.
%We conclude that the ratio $L_{\rm 1-0\,S(1)}$/$M_{\rm gas}$ traces
%the ratio of the number of PDRs to the number of molecular clouds.
%In ULIRGs, not only is the molecular gas mass higher, but this ratio
%is increased by 1--2 orders of magnitude, indicating that the star
%formation efficiency is also higher.
%This is the same conclusion as reached by other means \cite[see][and
%references therein]{san96}, and provides a natural explanation for the
%higher UV intensity on the PDRs.

%The situaton for the nuclei of other star forming galaxies is rather
%unclear.
%Some of the early work by \cite{pux88} suggested that the H$_2$ emission
%from nearby spiral galaxies with vigorous star formation was
%fluorescently excited.
%These authors measured 2-1\,S(1)/1-0\,S(1) ratios in the range
%0.44--0.86, consistent only with fluorescence in low density gas.
%However, many of the ratios are higher even than the theoretical
%values, and it may be that the results are biassed by difficulties of
%continuum levels and line blending due to the low resolution of the
%data.

\section{Spatial Variation of Line Flux}
\label{sec:spat}

In Section~\ref{sec:rat} we have discussed only the line emission in
the centre of the ULIRGs.
Figure~\ref{fig:spat} shows that in many cases it is extended beyond
this, and with a 1-0\,S(1)/Br$\gamma$ ratio which varies considerably
from that in the nucleus.
In Figure~\ref{fig:vel} we show position-velocity diagrams for the
objects, which show that (where it can be observed) the 1-0\,S(1) line
has the same velocity characteristics as the Br$\gamma$ line,
suggesting that the two lines arise from similar regions.
In contrast if, for example, the H$_2$ emission arose from material
blown out or shocked in a superwind 
\citep[e.g. as for NGC\,253,][]{sug03} 
one might expect to see broader or double-peaked profiles similar to
those observed in the H$\alpha$ lines by \cite{hec90}.
However, since the line intensity is very low in the circumnuclear
regions, we cannot rule out this possibility.

MCG\,$-$03$-$04$-$014 appears to be unique among these objects in
having the brightest Br$\gamma$ knots offset from the nucleus by
1--2\arcsec\ (0.7--1.4\,kpc).
This indicates that there is significant recent star formation in the
circumnuclear region.
At these locations the 1-0\,S(1)/Br$\gamma$ ratio has
decreased by a factor of 3 with respect to that in the nucleus,
perhaps indicative of lower PDR gas densities.

In many of the other ULIRGs, particularly IRAS\,01364$-$1042 and
IRAS\,14378$-$3651, there is evidence that the 1-0\,S(1)/Br$\gamma$
ratio increases in the circumnuclear region, although the surface
brightness diminishes rapidly.
In IRAS\,01364$-$1042 the ratio of 1-0\,S(1) to the continuum peaks at
radii of 0.5--1.0\arcsec\ (0.5--1.0\,kpc), while the Br$\gamma$
distribution is very centralised.
As in MCG\,$-$03$-$04$-$014, the reduction in 1-0\,S(1) surface
brightness could be due to lower gas densities in the PDRs.
IRAS\,16164$-$0746 appears to have very broad extended line emission
with little change in the 1-0\,S(1)/Br$\gamma$ ratio over at least
2\arcsec\ (1.1\,kpc).
The continuum emission is also broad, with a spatial FWHM of
1.4\arcsec\ (0.8\,kpc).
The image in Fig.~\ref{fig:acq} shows that the slit was
aligned close to the position angle of a prominent bar, and this may
be influencing the observed size of the nucleus.

Since the slit position angles for all the objects were assigned
rather abitrarily (due to the quality of previously available
images) and in most cases do not lie along either the major or the
minor axis of the galaxy, it is difficult to draw strong conclusions
about the extended emission.
However, it is clear that the characteristics of the emission lines
(strength and ratios) change in the circumnuclear region, and hence
the excitation conditions must also change.

\section{Conclusions}
\label{sec:conc}

We have presented K-band spectra of 7 ULIRGs.
Fitting stellar templates to line-free parts of the spectra has
allowed us to fit and remove the absorption features in order to
improve our detection limit for faint lines.
We have measured line fluxes of Br$\gamma$ and H$_2$ lines from the
$\nu=1$, 2, and 3 levels, and conclude the following:

(1) The nuclear H$_2$ line ratios show very little variation between
the objects, suggesting that the excitation mechanism and conditions
are similar in all cases, at least in the nuclei.
However, the 1-0\,S(1) and Br$\gamma$ line ratio and equivalent widths
vary away from the nucleus, suggesting that in the circumnuclear
region the excitation mechanisms may be different, perhaps as a result
of a different evolutionary phase of the star formation.

(2) We find that neither shocks nor models of low
density PDRs can satisfactorily fit the nuclear H$_2$ line ratios;
but FUV-pumped gas in high density PDRs can account for the observed
ratios and fluxes.
Parameter ranges that produce PDR models consistent with the
observed H$_2$ line ratios are: gas densities $n_{\rm
H}=10^{4-5}$\,cm$^{-3}$, temperatures at the outer edge of the clouds
$T\approx1000$\,K, and illuminating FUV fields
$\chi = 10^3$--$10^5$ times more intense than the local interstellar
field.

(3) We have constructed a simple model of star formation based on
fluorescence of gas in PDRs at the edge of H{\small \,II} regions
surrounding compact star clusters.
This model can account for the ratios of the He{\small \,I}, Br$\gamma$,
and fluorescent H$_2$ line intensities, and implies
young (1--5\,Myr) clusters containing massive (120 M$_\odot$) stars.
Our model implies that the nuclear K-band continuum is dominated
by a separate older stellar population.

(4) The PDRs in ULIRGs and those in the inner few hundred parsecs of
the Galaxy have similar 1-0\,S(1) surface brightnesses.
Since the 1-0\,S(1) emission in both cases is predominantly from PDRs,
the quantity $L_{\rm 1-0\,S(1)}$/$M_{\rm gas}$ traces the ratio of the
number of PDRs to the number of molecular clouds -- i.e. the star
formation efficiency.
This ratio is 10--100 times larger in ULIRGs than in the Galactic
Center.

%----------------------------------------------------------------------

\acknowledgments

The authors are grateful to the staff at the Paranal Observatory for
carrying out in service mode the observations presented in this paper.
We thank the German-Israeli Foundation (grant I-0551-186.07/97)
for support. 
We thank R. Genzel, D. Lutz, and the referee
for comments and helpful suggestions.

%----------------------------------------------------------------------

%----------------------------------------------------------------------

\clearpage

%----------------------------------------------------------------------

\begin{deluxetable}{lrrrrrrr}

\tabletypesize{\small}
\tablecaption{Basic Data for the Galaxies Observed\label{tab:basdat}}
\tablehead{

\colhead{object} & 
\colhead{RA} & 
\colhead{dec} & 
\colhead{cz} & 
\colhead{S$_{12}$} & 
\colhead{S$_{25}$} & 
\colhead{S$_{60}$} & 
\colhead{S$_{100}$} \\

\colhead{} & 
\colhead{(J2000)} & 
\colhead{(J2000)} & 
\colhead{(km\,s$^{-1}$)} & 
\colhead{(Jy)} & 
\colhead{(Jy)} & 
\colhead{(Jy)} & 
\colhead{(Jy)} \\

}
\startdata

MCG\,$-$03$-$04$-$014 & 01 10 08.9 & $-$16 51 10 & 10040 & 0.301 &
0.846 & 6.48 & 10.44 \\
IRAS\,01364$-$1042    & 01 38 52.9 & $-$10 27 12 & 14520 & $<$0.077 &
0.396 & 6.16 & 6.70 \\
MCG\,$-$03$-$12$-$002\tablenotemark{a} & 04 21 20.0 & $-$18 48 39 &  9652 & 0.185 &
0.430 & 5.75 & 8.20 \\
IRAS\,14378$-$3651    & 14 40 59.4 & $-$37 04 33 & 20277 & $<$0.116 &
0.525 & 6.19 & 6.34 \\
IRAS\,16164$-$0746    & 16 19 11.9 & $-$07 54 03 &  8140 & $<$0.350 &
0.562 & 10.20 & 13.72 \\
IRAS\,19458$+$0944    & 19 48 15.5 & $+$09 52 02 & 29964 & $<$0.250 &
$<$0.278 & 3.95 & 7.11 \\
IRAS\,20414$-$1651\tablenotemark{b}    & 20 44 17.4 & $-$16 40 14 & 26107 & $<$0.647 & 
0.346 & 4.36 & 5.25 \\

\enddata

\tablenotetext{a}{MCG\,$-$03$-$12$-$002 is a pair with a separation of
17\arcsec; we observed only the northern component, to which the IRAS
data is closer.}

\tablenotetext{b}{IRAS\,20414$-$1651 has two nuclei separated by
2.6\arcsec, of which only one is visible in the K-band.}

\tablecomments{IRAS data is from the Faint Source Catalogue and Point
Source Catalogue}

\end{deluxetable}

%----------------------------------------------------------------------

\begin{deluxetable}{lcccc}

\tablecaption{Measured Data\label{tab:measdat}}
\tablehead{

\colhead{object} & 
\colhead{Distance\tablenotemark{a}} & 
\colhead{K-mag} & 
\colhead{$F_{S(1)}$\tablenotemark{b}} & 
\colhead{$\log{\frac{L_{S(1)}}{L_\odot}}$\tablenotemark{b}} \\

\colhead{} & 
\colhead{(Mpc)} & 
\colhead{} & 
\colhead{10$^{-18}$\,W\,m$^{-2}$} & 
\colhead{} \\

}
\startdata

MCG\,$-$03$-$04$-$014 & 145 & 13.4 & 2.85 & 6.27 \\
IRAS\,01364$-$1042    & 210 & 13.4 & 2.03 & 6.45 \\
MCG\,$-$03$-$12$-$002 & 139 & 15.0 & 0.53 & 5.51 \\
IRAS\,14378$-$3651    & 294 & 14.7 & 0.93 & 6.40 \\
IRAS\,16164$-$0746    & 117 & 13.2 & 1.20 & 5.71 \\
IRAS\,19458$+$0944    & 438 & 14.4 & 1.90 & 7.06 \\
IRAS\,20414$-$1651    & 381 & 13.8 & 1.01 & 6.66 \\

\enddata

\tablenotetext{a}{Luminosity distance calculated using
$H_0=70$\,km\,s$^{-1}$\,Mpc$^{-1}$ and $q_0=0.5$.}

\tablenotetext{b}{1-0\,S(1) line flux extracted in a 0.75\arcsec\ length of a
1.00\arcsec\ slit; errors are given with the line ratios in
Table~\ref{tab:ratios}.}

\end{deluxetable}

%----------------------------------------------------------------------

\begin{deluxetable}{lrrrrr}

\tabletypesize{\small}
\tablecaption{Parameters for the H$_2$ PDR Models\label{tab:models}}
\tablehead{

\colhead{model} & 
\colhead{$\chi$} & 
\colhead{$n_{\rm H}$} & 
\colhead{$T_{\rm max}$\tablenotemark{a}} & 
\colhead{$I_{\rm 1-0S(1)}$\tablenotemark{b}} &
\colhead{$I_{\rm H_2}$\tablenotemark{c}} \\

\colhead{} & 
\colhead{} & 
\colhead{(cm$^{-3}$)} & 
\colhead{(K)} & 
\colhead{(W\,m$^{-2}$\,sr$^{-1}$)} &
\colhead{(W\,m$^{-2}$\,sr$^{-1}$)} \\

}
\startdata

1 & $10^2$ & $10^3$ & $10^2$ & $1.29\times10^{-9}$ & $7.06\times10^{-8}$ \\
2 & $10^3$ & $10^4$ & $10^3$ & $5.17\times10^{-8}$ & $2.88\times10^{-6}$ \\
3 & $10^3$ & $10^4$ & $2\times10^3$ & $2.00\times10^{-6}$ & $3.22\times10^{-5}$ \\
4 & $10^5$ & $10^4$ & $10^3$ & $7.41\times10^{-8}$ & $4.04\times10^{-6}$ \\
5 & $10^5$ & $10^6$ & $10^3$ & $8.82\times10^{-7}$ & $1.80\times10^{-5}$ \\

\enddata

\tablenotetext{a}{For model 1 $T=100$\,K at all cloud depths. 
For the other models, the temperature varies as given in
Eq.~\ref{eq:T}.}

\tablenotetext{b}{The 1-0\,S(1) intensity is that predicted by the
model.
Intensities for other lines can be found by using the model line
ratios given in Table~\ref{tab:ratios}.}

\tablenotetext{c}{The total H$_2$ intensity summed over all the H$_2$ lines
is that predicted by the model.}

\end{deluxetable}

%----------------------------------------------------------------------

\begin{deluxetable}{lrrrrrrrrrrr}

\tabletypesize{\tiny}
\tablecaption{Relative Line Fluxes\label{tab:ratios}}
\tablehead{

\colhead{} &
\colhead{} & 
\multicolumn{10}{c}{line and wavelength ($\mu$m)} \\

\colhead{object/model} & 
\colhead{1$\sigma$\tablenotemark{a}} & 
\colhead{1-0S(2)} & 
\colhead{He\,I} & 
\colhead{3-2S(5)} & 
\colhead{2-1S(3)} & 
\colhead{1-0S(1)} & 
\colhead{2-1S(2)} & 
\colhead{Br$\gamma$} & 
\colhead{3-2S(3)} & 
\colhead{1-0S(0)} & 
\colhead{2-1S(1)} \\

\colhead{} &
\colhead{} &
\colhead{2.0338} &
\colhead{2.0587} &
\colhead{2.0656} & 
\colhead{2.0735} & 
\colhead{2.1218} & 
\colhead{2.1542} & 
\colhead{2.1661} &
\colhead{2.2014} & 
\colhead{2.2233} & 
\colhead{2.2477} \\

}
\startdata

MCG\,$-$03$-$04$-$014 & 0.053 & 0.354 & 0.316 & $<$0.160 & 0.128 &
1.000 & $<$0.160 & 1.254 & $<$0.160 & 0.298 & 0.194 \\

IRAS\,01364$-$1042    &0.016 & 0.365 & 0.282 & $<$0.048 & 0.139 &
1.000 & 0.066 & 0.600 & 0.062 & 0.262 & 0.140 \\

MCG\,$-$03$-$12$-$002 & 0.042 & 0.335 & 0.221 & $<$0.126 & 0.173 &
1.000 & 0.078 & 0.657 & 0.088 & 0.218 & 0.113 \\

IRAS\,14378$-$3651    & 0.034 & 0.325 & 0.459 & $<$0.103 & 0.144 &
1.000 & $<$0.103 & 1.153 & $<$0.103 & 0.268 & --- \\

IRAS\,16164$-$0746    & 0.022 & --- & 0.665 & $<$0.065 & 0.169 & 
1.000 & 0.078 & 1.331 & 0.079 & 0.321 & 0.138 \\

IRAS\,19458$+$0944    & 0.052 & 0.317 & 0.386 & $<$0.157 & 0.155 &
1.000 & $<$0.157 & 1.038 & --- & --- & --- \\

IRAS\,20414$-$1651    & 0.061 & 0.328 & 0.348 & $<$0.183 & 0.113 &
1.000 & $<$0.183 & 0.797 & $<$0.183 & 0.258 & 0.123 \smallskip\\

mean H$_2$ ratios & & 0.337 & & $<$0.120 & 0.146 & 
1.000 & 0.074 & & 0.076 & 0.271 & 0.142 \\

std. dev. of ratios & & 0.018 & & & 0.022 & 
 & 0.007 & & 0.013 & 0.035 & 0.031 \\

\\

thermal:\ 1000\,K & & 0.27\phn & & 0.00\phn & 0.00\phn & 
1.00\phn & 0.00\phn & & 0.00\phn & 0.27\phn & 0.01\phn \\

\phm{thermal:\ }2000\,K & & 0.37\phn & & 0.00\phn & 0.08\phn & 
1.00\phn & 0.03\phn & & 0.01\phn & 0.21\phn & 0.08\phn \\

%Bla96:
%UV, $10^3$\,cm$^{-3}$ \tablenotemark{b} 
%& & 0.50\phn & & 0.07\phn & 0.35\phn & 
%1.00\phn & 0.28\phn & & 0.18\phn & 0.46\phn & 0.56\phn \\

PDR\tablenotemark{b}\ \ :\ model 1 &
& 0.49\phn & & 0.02\phn & 0.22\phn & 
1.00\phn & 0.25\phn & & 0.10\phn & 0.54\phn & 0.53\phn \\

\phm{UV\tablenotemark{b}\ \ :\ }model 2 &
& 0.32\phn & & 0.04\phn & 0.15\phn & 
1.00\phn & 0.08\phn & & 0.07\phn & 0.33\phn & 0.17\phn \\

\phm{UV\tablenotemark{b}\ \ :\ }model 3 &
& 0.35\phn & & 0.00\phn & 0.05\phn & 
1.00\phn & 0.02\phn & & 0.00\phn & 0.22\phn & 0.05\phn \\

\phm{UV\tablenotemark{b}\ \ :\ }model 4 & 
& 0.31\phn & & 0.05\phn & 0.16\phn & 
1.00\phn & 0.08\phn & & 0.07\phn & 0.32\phn & 0.17\phn \\

\phm{UV\tablenotemark{b}\ \ :\ }model 5 & 
& 0.29\phn & & 0.02\phn & 0.10\phn & 
1.00\phn & 0.04\phn & & 0.04\phn & 0.33\phn & 0.11\phn \\

\enddata

\tablenotetext{a}{The 1$\sigma$ error applies to all line fluxes
given, including 1-0\,S(1).
It is measured directly from the spectrum as the {\small rms} of the
residual after subtraction of both the stellar continuum and emission
lines.}

\tablenotetext{b}{For details of the PDR models, see
Table~\ref{tab:models}.}

%\tablenotetext{c}{Models Mv3o and Rh3o from \citet{dra96} with
%$n=10^5$\,cm$^{-3}$ and $n=10^6$\,cm$^{-3}$ respectively, and a ratio
%of UV field to density $\chi/n=0.1$.}

\tablecomments{Lines with no ratio given were not within the
wavelength covered; 3$\sigma$ upper limits are given for lines within
this range that were not detected.}

\end{deluxetable}

%----------------------------------------------------------------------

\begin{deluxetable}{lccccc}

\tablecaption{Derived Data\label{tab:derdat}}
\tablehead{

\colhead{object} & 
\colhead{$\log{\frac{L_{bol}}{L_\odot}}$\tablenotemark{a}} & 
\colhead{$\log{\frac{L_{\rm K}}{L_\odot}}$} & 
\colhead{$\log{\frac{Q_{\rm Lyc}}{\rm ph\,s^{-1}}}$} & 
\colhead{$\frac{Q_{\rm H_2}}{Q_{\rm Lyc}}$\tablenotemark{b}} &
\colhead{$100\times\frac{Q^{\rm min}_{\rm He}}{Q_{\rm Lyc}}$\tablenotemark{b}} \\

}
\startdata

MCG\,$-$03$-$04$-$014 & 11.62 & 8.83 & 53.82 & $0.33\pm0.06$ & $1.92\pm0.06$ \\
IRAS\,01364$-$1042    & 11.79 & 9.15 & 53.68 & $0.75\pm0.01$ & $3.58\pm0.01$ \\
MCG\,$-$03$-$12$-$002 & 11.48 & 8.15 & 52.78 & $0.86\pm0.05$ & $2.56\pm0.10$ \\
IRAS\,14378$-$3651    & 12.09 & 8.93 & 53.91 & $0.41\pm0.02$ & $3.03\pm0.02$ \\
IRAS\,16164$-$0746    & 11.52 & 8.72 & 53.29 & $0.41\pm0.01$ & $3.80\pm0.01$ \\
IRAS\,19458$+$0944    & 12.27 & 9.39 & 54.53 & $0.48\pm0.06$ & $2.83\pm0.06$ \\
IRAS\,20414$-$1651    & 12.18 & 9.51 & 54.02 & $0.46\pm0.14$ & $3.33\pm0.12$ \\

\enddata

\tablenotetext{a}{$L_{bol}$ is calculated from the IRAS fluxes for the
range 8--1000\,$\mu$m, using the formula given in \citet{tok00}.}

\tablenotetext{b}{$Q_{\rm H_2}$ is calculated from a
fluorescently excited H$_2$ line for PDR model 2.
Quoted errors include uncertainties in both rates.}

\end{deluxetable}

%----------------------------------------------------------------------

\begin{deluxetable}{cccccc}

\tablecaption{Star Formation Models\label{tab:Qratio}}
\tablehead{

\colhead{age (Myr)} & 
\colhead{$\log{\frac{L_{\rm K}}{L_\odot}}$} & 
\colhead{$\log{\frac{Q_{\rm Lyc}}{\rm ph\,s^{-1}}}$} & 
\colhead{$\frac{Q_{\rm H_2}}{Q_{\rm Lyc}}$} &
\colhead{$100\times\frac{Q_{\rm He}}{Q_{\rm Lyc}}$} &
\colhead{$\log{\frac{L_{\rm bol}}{L_\odot}}$} \\

}
\startdata
%AS I have replaced ``initial mass'' with ``cluster mass''
\multicolumn{6}{l}{Instantaneous, cluster mass $10^5$\,M$_\odot$,
Salpeter IMF, $M_{\rm upper} = 120$\,M$_\odot$} \\
\phn\phn1 & 5.48 & 51.95 & \phn0.63 &    19.02 & \phn8.27 \\
\phn\phn5 & 5.06 & 50.78 & \phn5.12 & \phn1.51 & \phn7.89 \\
   \phn10 & 5.70 & 49.24 &    61.85 & \phn0.28 & \phn7.48\smallskip\\

\multicolumn{6}{l}{Instantaneous, cluster mass $10^5$\,M$_\odot$,
Salpeter IMF, $M_{\rm upper} = 30$\,M$_\odot$} \\
\phn\phn1 & 4.66 & 50.91 & \phn2.25 & \phn9.41 & \phn7.67 \\
\phn\phn5 & 4.63 & 50.80 & \phn5.57 & \phn1.05 & \phn7.80 \\
   \phn10 & 5.76 & 49.31 &    61.88 & \phn0.28 & \phn7.55\smallskip\\

\multicolumn{6}{l}{Continuous, $SFR=1$\,M$_\odot$\,yr$^{-1}$,
Salpeter IMF, $M_{\rm upper} = 120$\,M$_\odot$} \\
\phn\phn1 & 6.48 & 52.95 & \phn0.61 &    20.54 & \phn9.25 \\
\phn\phn3 & 6.89 & 53.37 & \phn0.80 &    16.36 & \phn9.76 \\
\phn\phn5 & 7.03 & 53.42 & \phn1.21 &    15.10 & \phn9.90 \\
   \phn10 & 7.55 & 53.43 & \phn1.34 &    14.80 &    10.03 \\
      100 & 8.06 & 53.43 & \phn2.02 &    14.81 &    10.20\smallskip\\

\multicolumn{6}{l}{Continuous, $SFR=1$\,M$_\odot$\,yr$^{-1}$,
Salpeter IMF, $M_{\rm upper} = 30$\,M$_\odot$} \\
\phn\phn1 & 5.66 & 51.91 & \phn2.28 & \phn9.42 & \phn8.66 \\
\phn\phn3 & 6.15 & 52.40 & \phn2.89 & \phn8.50 & \phn9.18 \\
\phn\phn5 & 6.37 & 52.60 & \phn3.81 & \phn6.00 & \phn9.44 \\
   \phn10 & 7.48 & 52.66 & \phn5.36 & \phn5.36 & \phn9.74 \\
      100 & 8.09 & 52.67 & \phn8.66 & \phn5.22 &    10.07\smallskip\\

\enddata

\end{deluxetable}

%----------------------------------------------------------------------

\clearpage

%----------------------------------------------------------------------

%----------------------------------------------------------------------

\begin{figure}
\centerline{\psfig{file=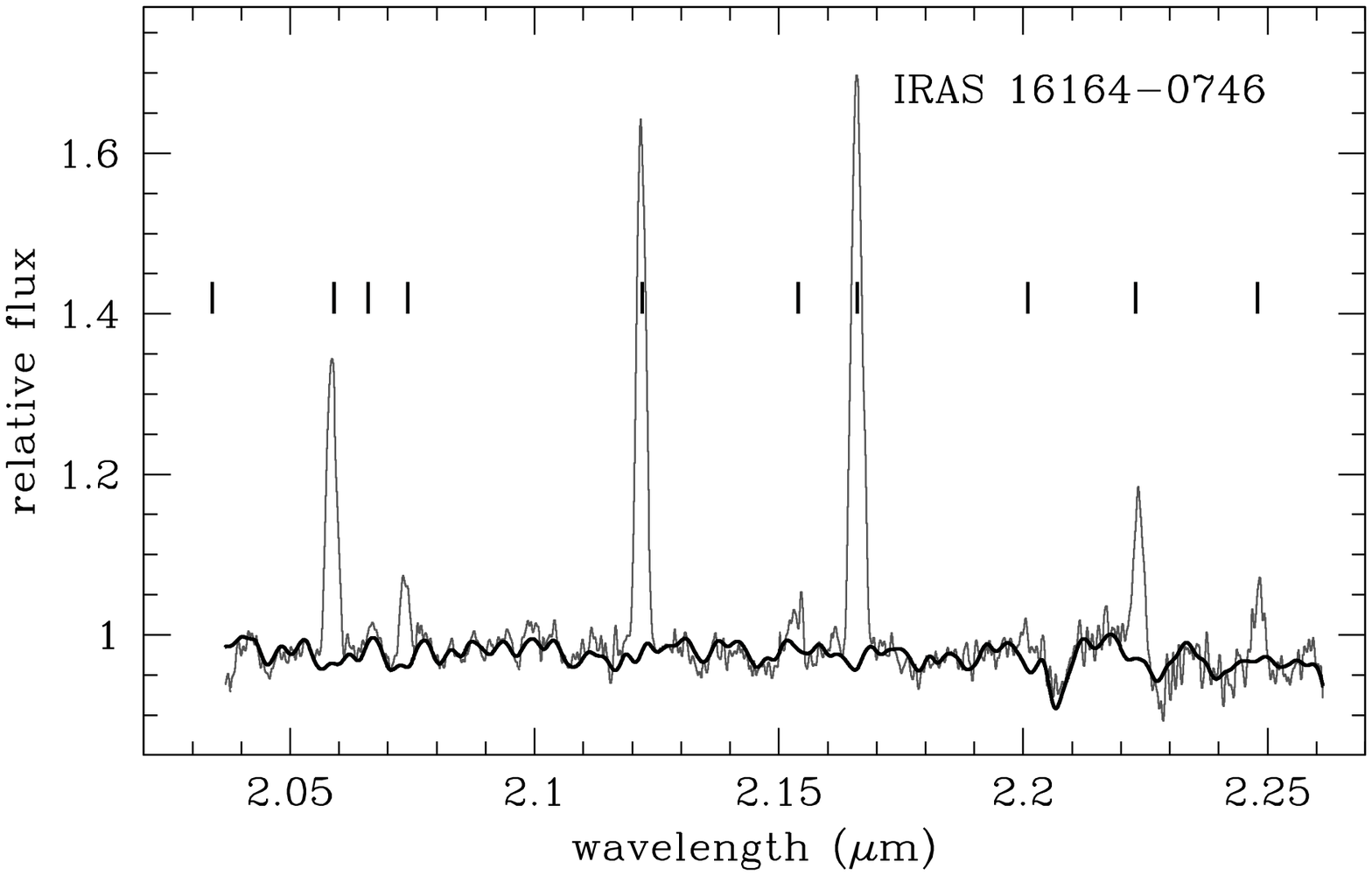,width=8cm}\hspace{5mm}\psfig{file=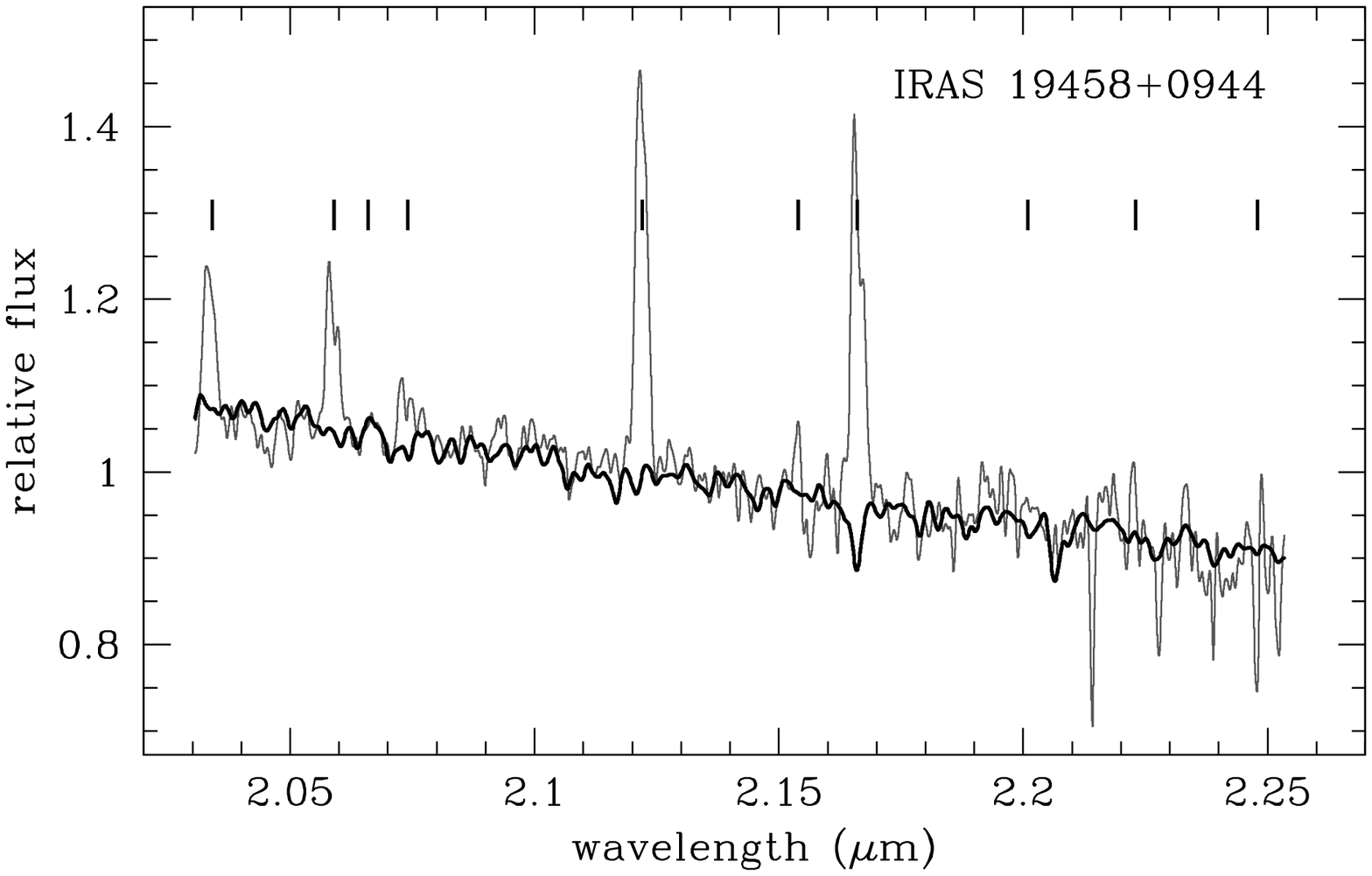,width=8cm}}
\vspace{5mm}
\centerline{\psfig{file=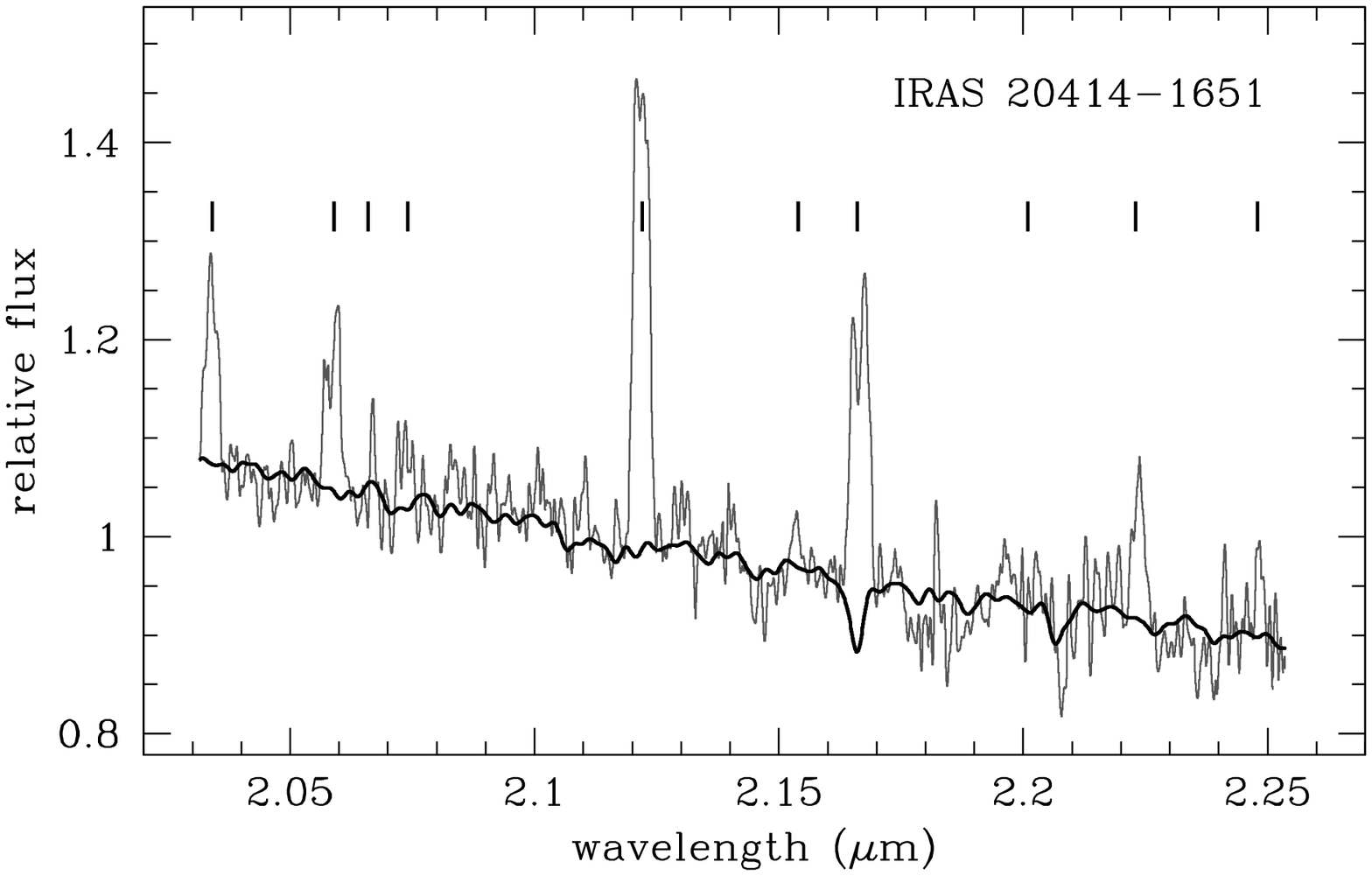,width=8cm}}
%\caption{As for Fig.~\ref{fig:spec}.}
%\label{fig:specb}
\end{figure}

%----------------------------------------------------------------------

\begin{figure}
\centerline{\psfig{file=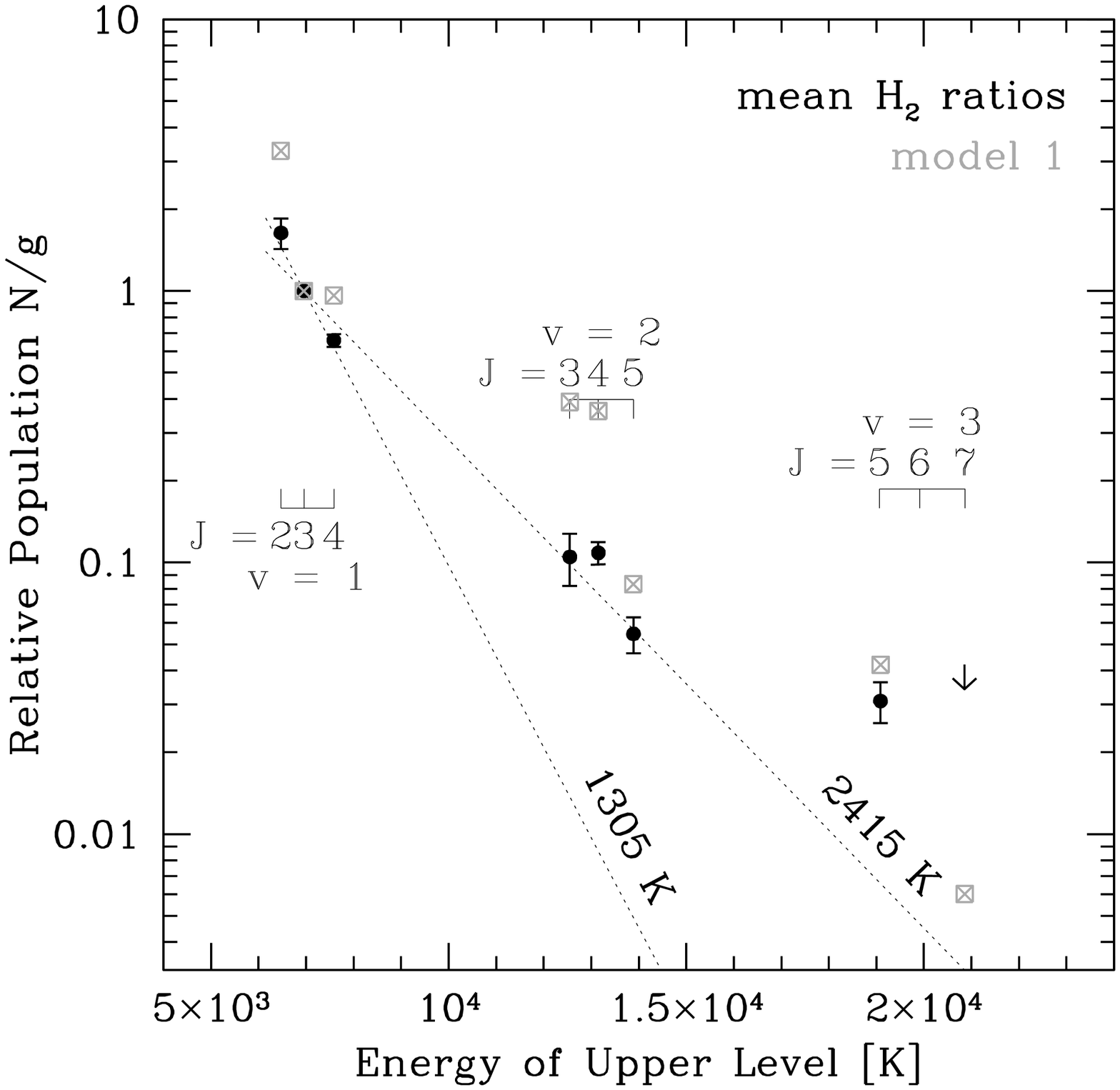,width=7cm}\hspace{5mm}\psfig{file=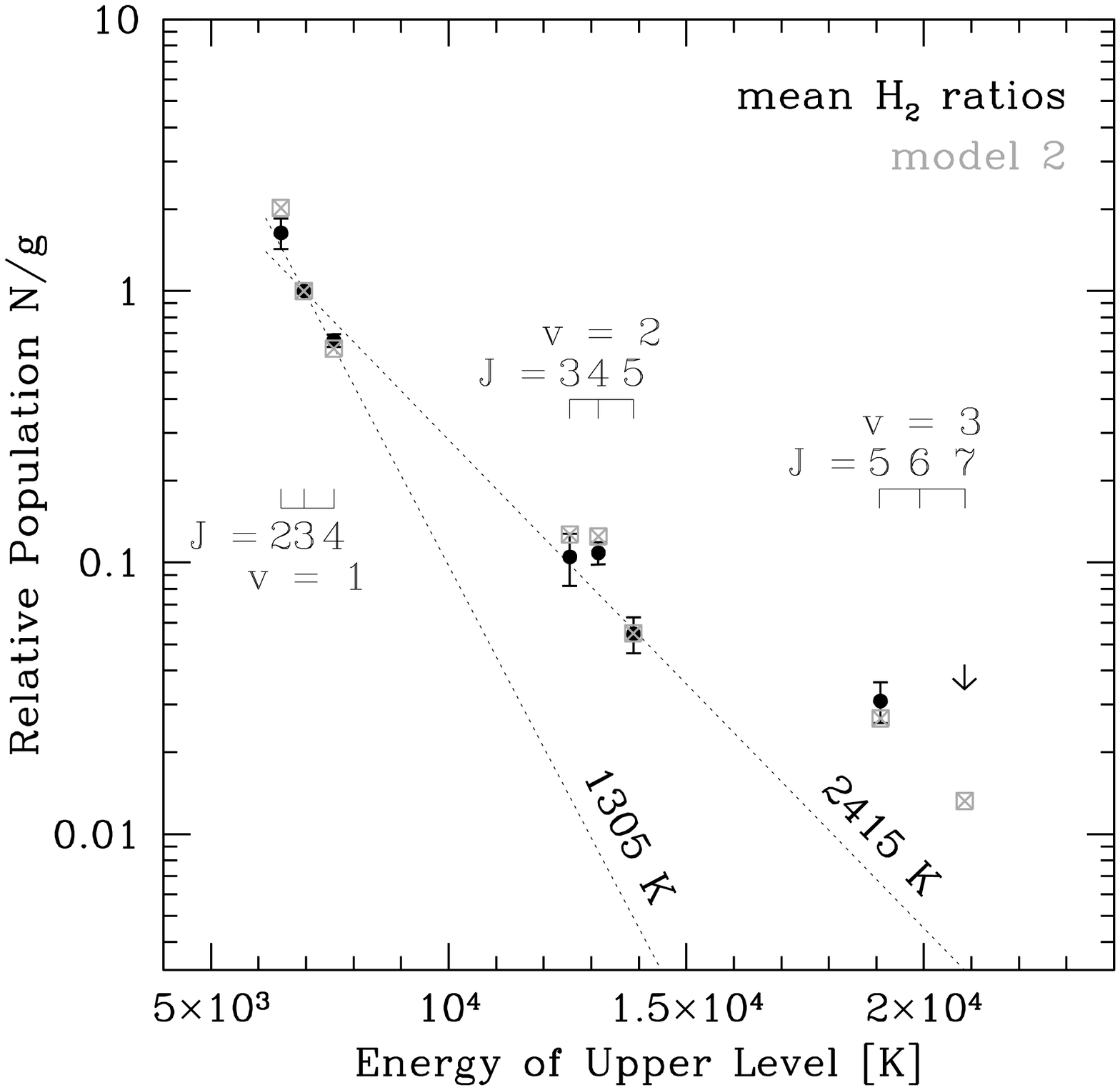,width=7cm}}
\vspace{5mm}
\centerline{\psfig{file=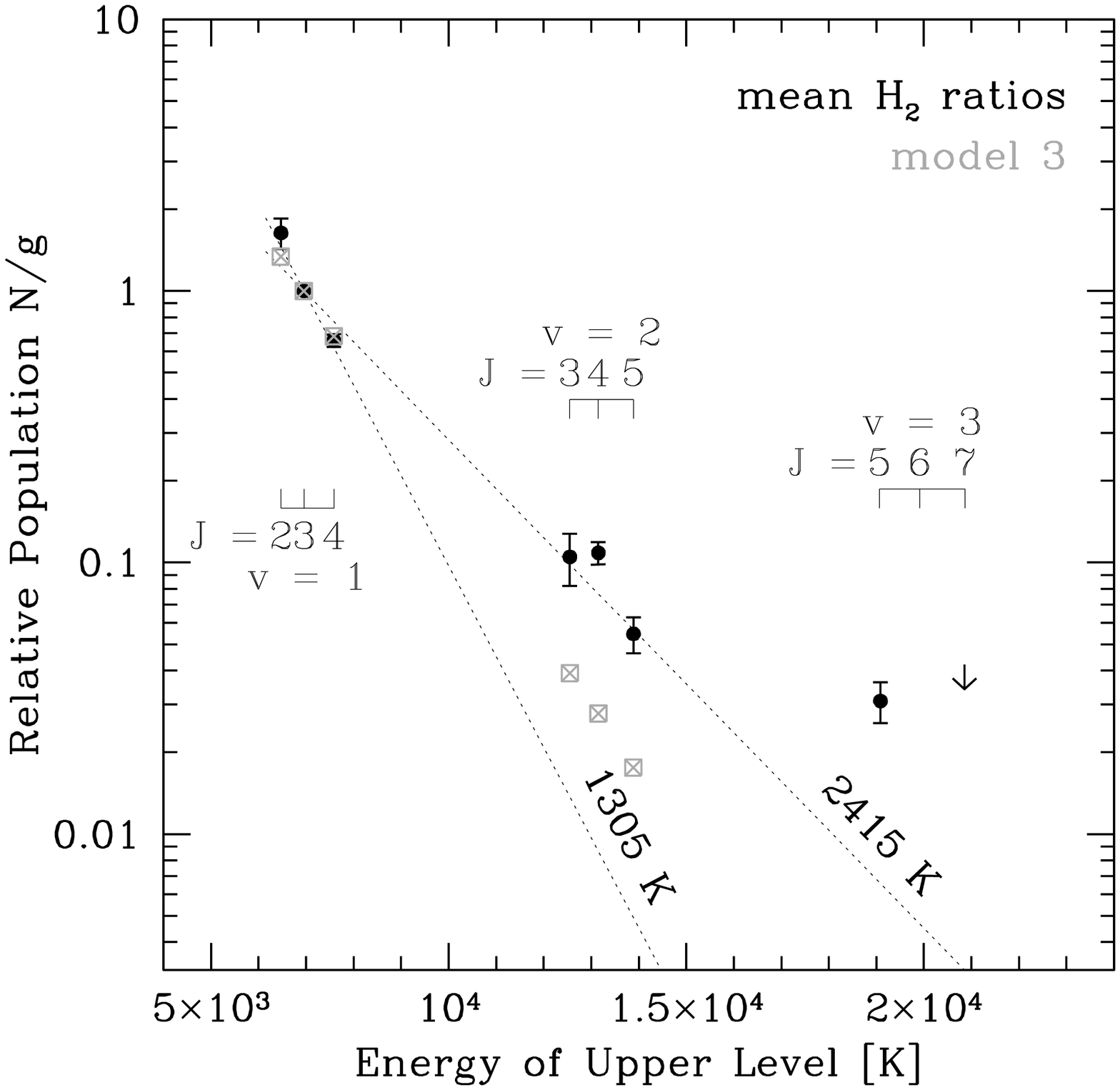,width=7cm}\hspace{5mm}\psfig{file=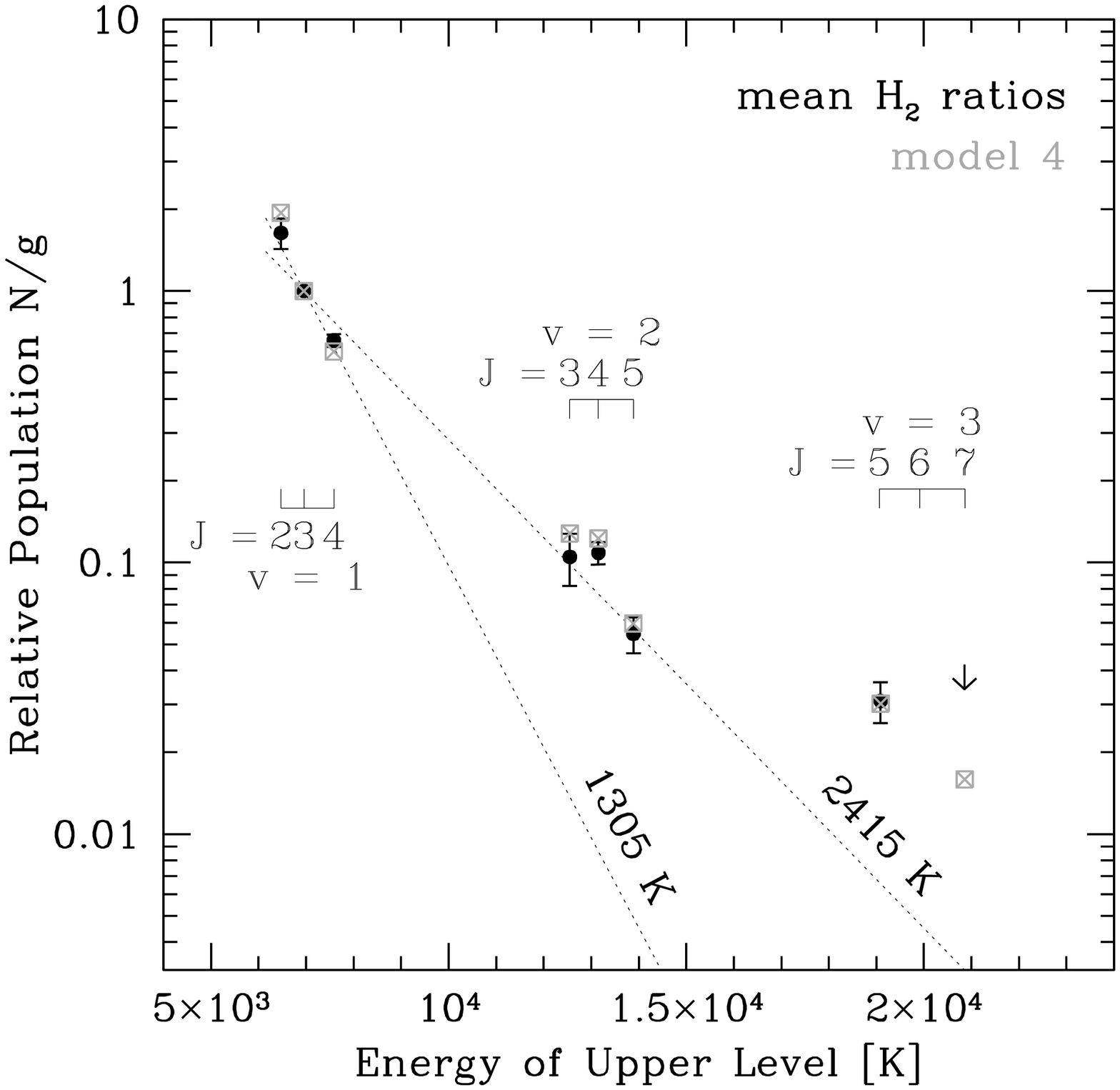,width=7cm}}
\centerline{\psfig{file=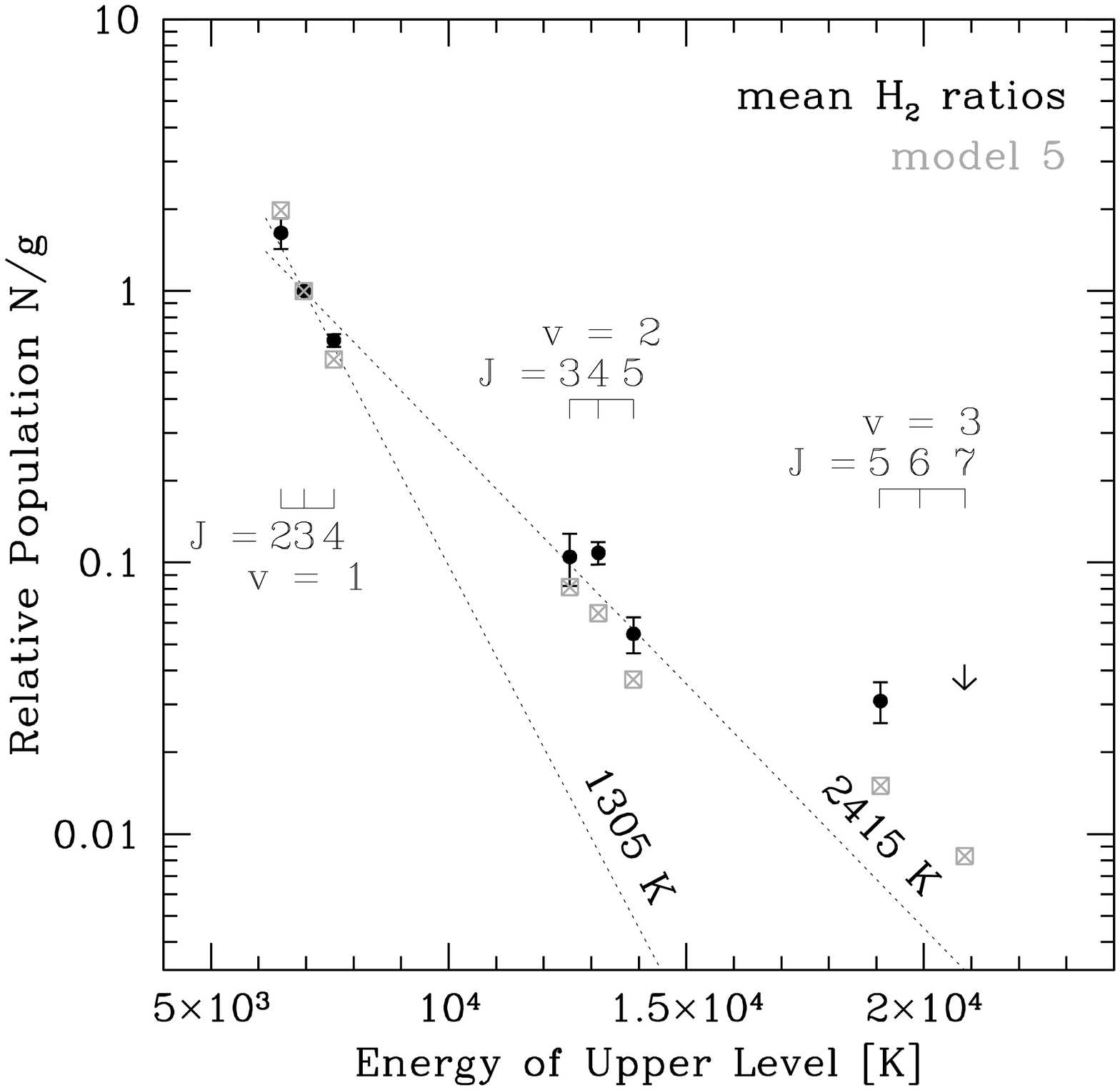,width=7cm}\hspace{5mm}\hspace{7cm}}
\caption{Excitation diagrams for the mean line ratios, and the 5 PDR
models discussed in the text.
Errors denote the standard deviation in the values used to calculate
the mean; arrows denote mean of the $3\sigma$ upper limits.
The two dotted lines indicate where the points for 
the best fitting purely thermal single temperature models would
lie; the fits 
are to all the data, and also to that in the $\nu=1$ level only.
The grey points show the populations for our various PDR models.}
\label{fig:popmean}
\end{figure}

%----------------------------------------------------------------------

\begin{figure}
\centerline{\psfig{file=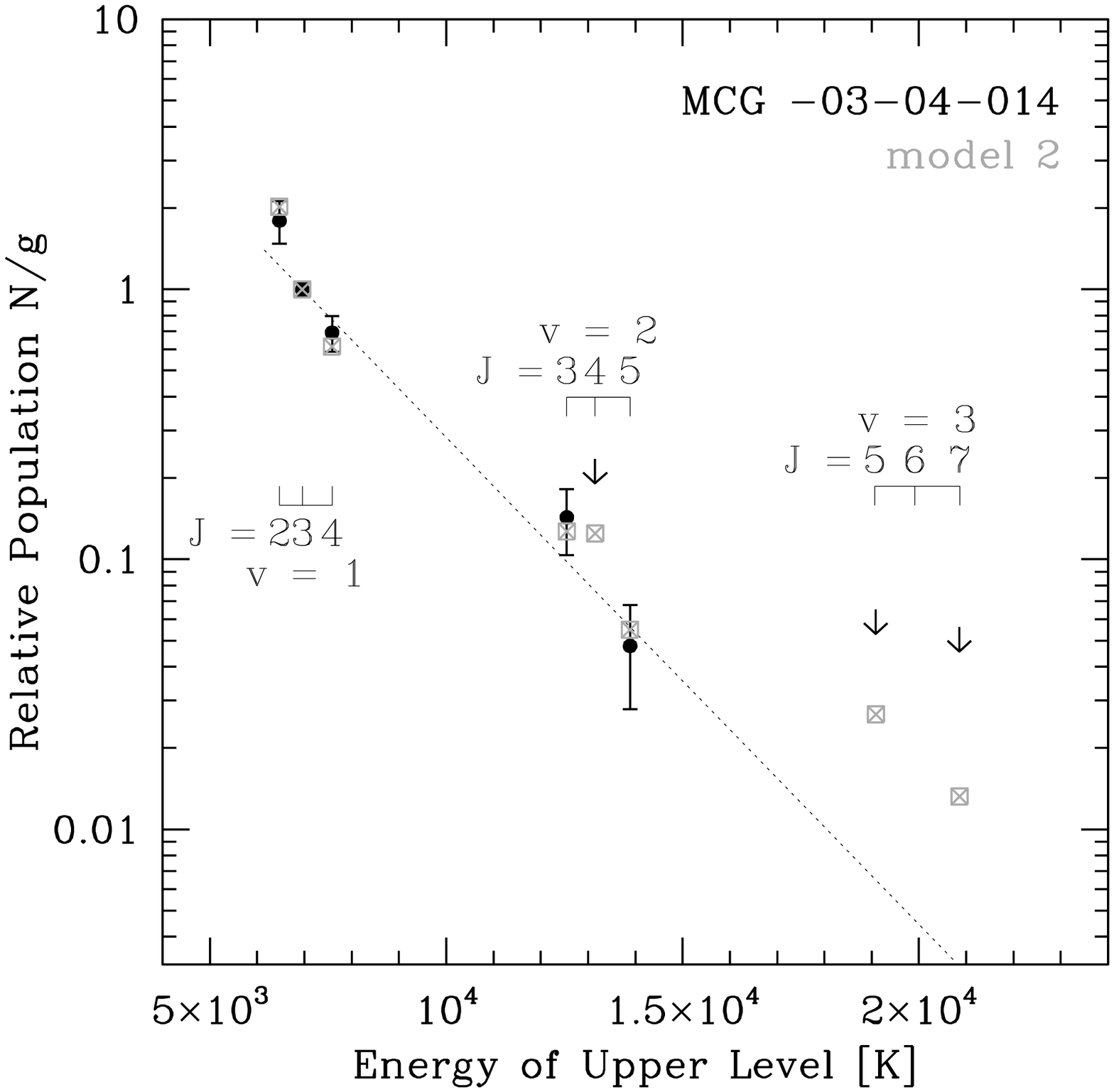,width=8cm}\hspace{5mm}\psfig{file=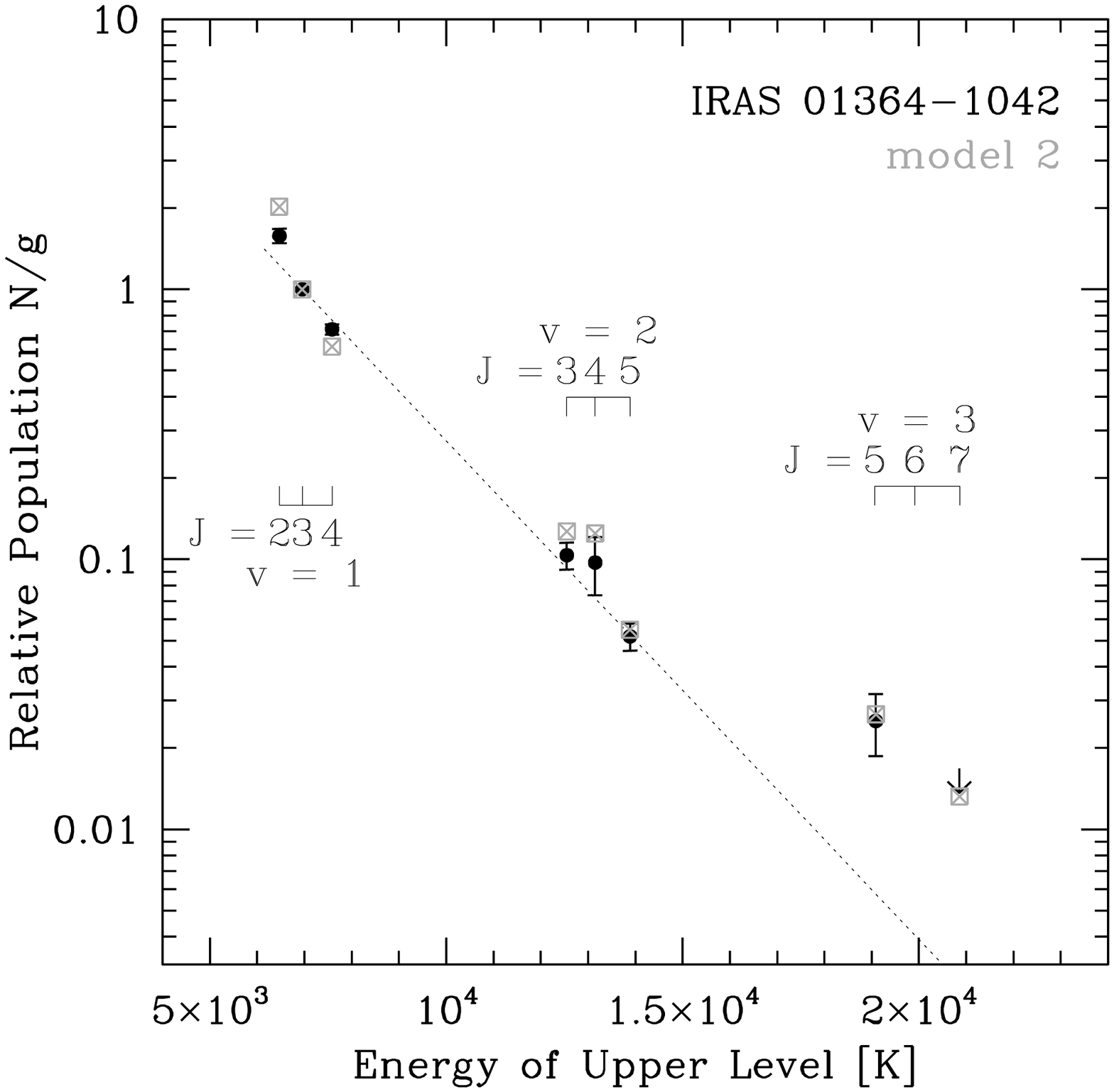,width=8cm}}
\vspace{5mm}
\centerline{\psfig{file=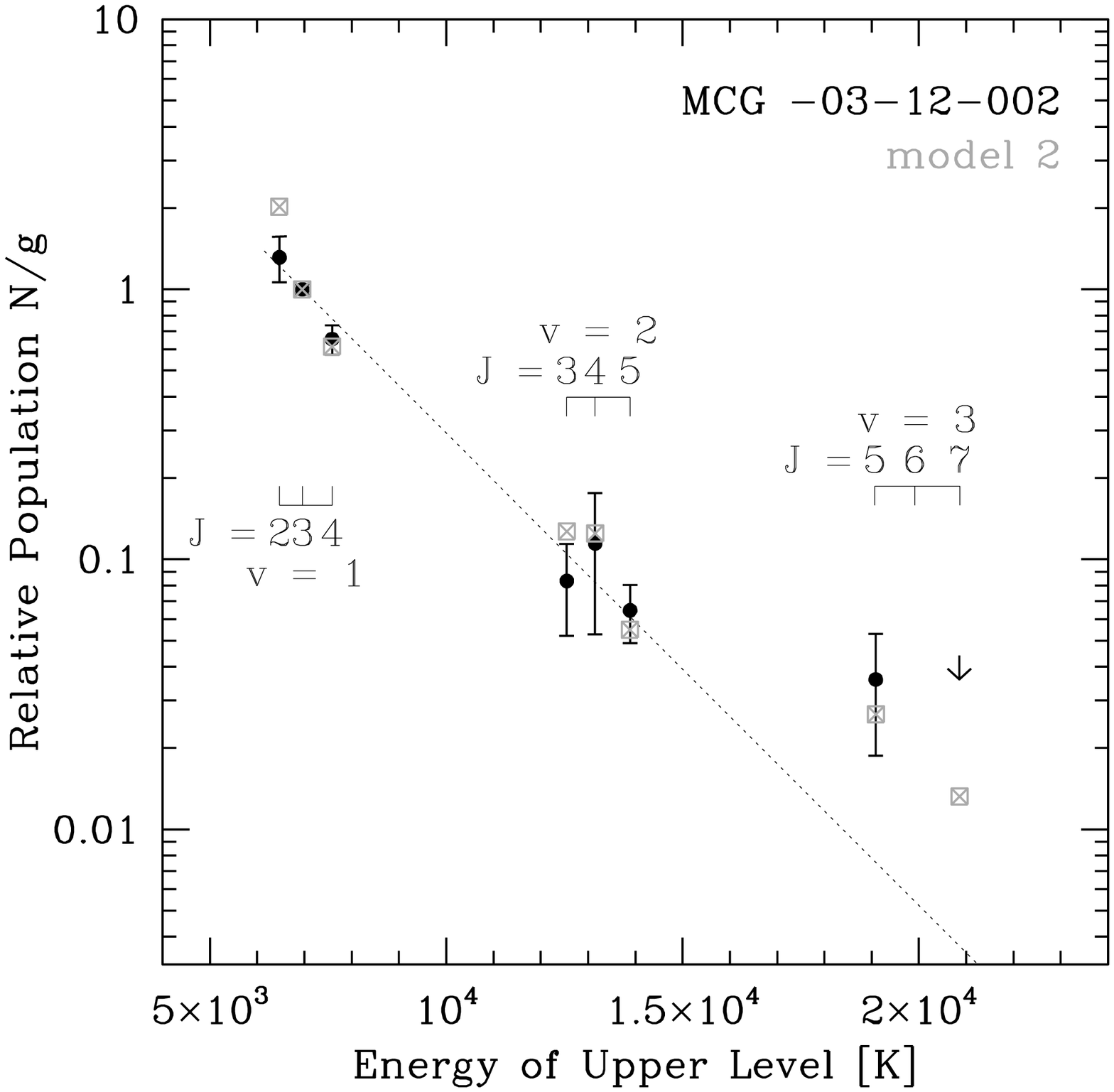,width=8cm}\hspace{5mm}\psfig{file=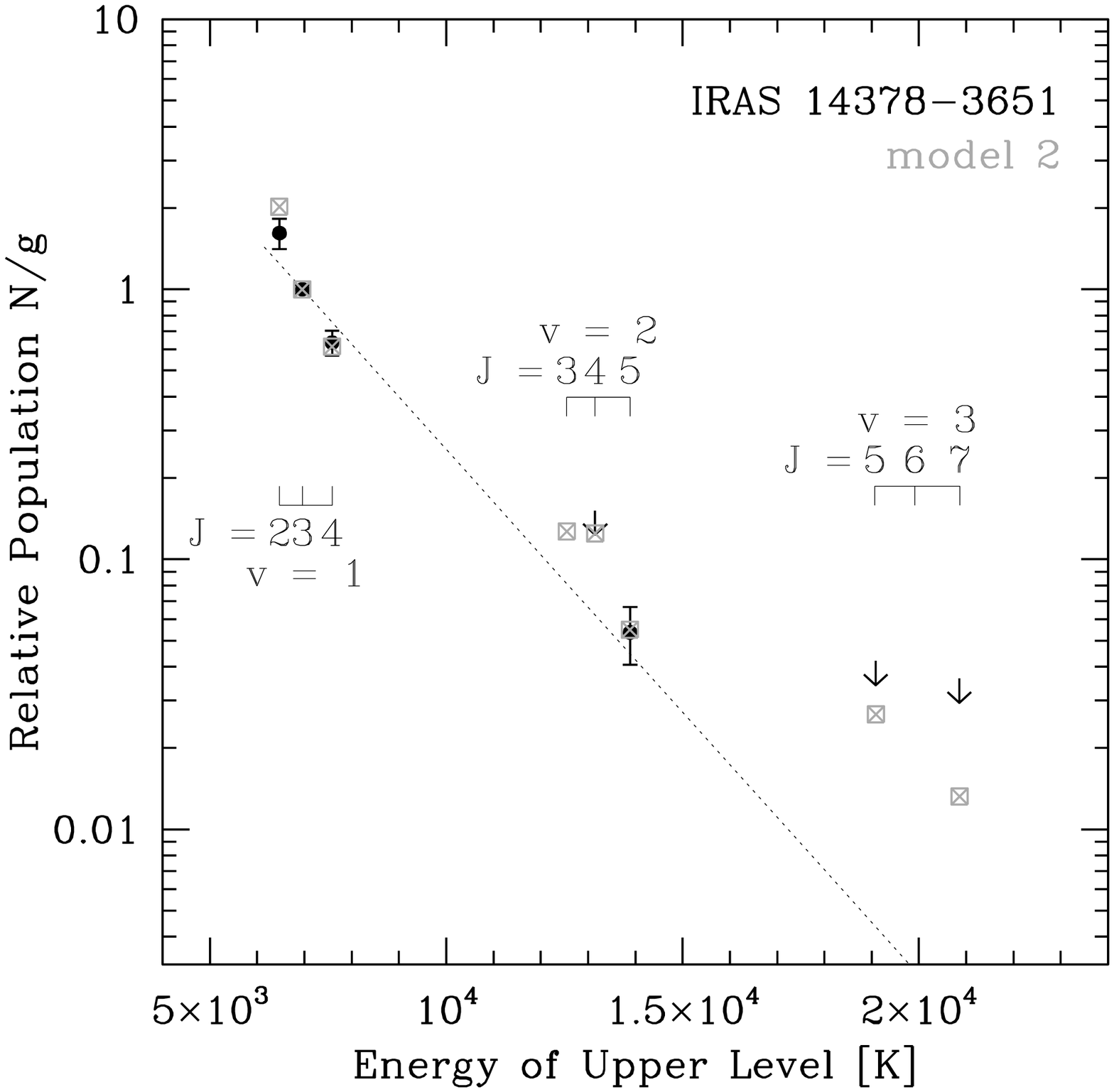,width=8cm}}
\caption{Excitation diagrams for the hot hydrogen molecules in each
object.
Arrows denote $3\sigma$ upper limits, derived from the residual
spectrum after subtraction of the stellar continuum and line emission.
The dotted lines indicate where the points for the best fitting purely
thermal single temperature model would lie.
%However, in a number of cases there are clear discrepancies between
%the model and the data, indicating that this cannot explain the data.
The overlaid grey points show the populations for our PDR model 2.}
\label{fig:pop}
\end{figure}

\begin{figure}
\centerline{\psfig{file=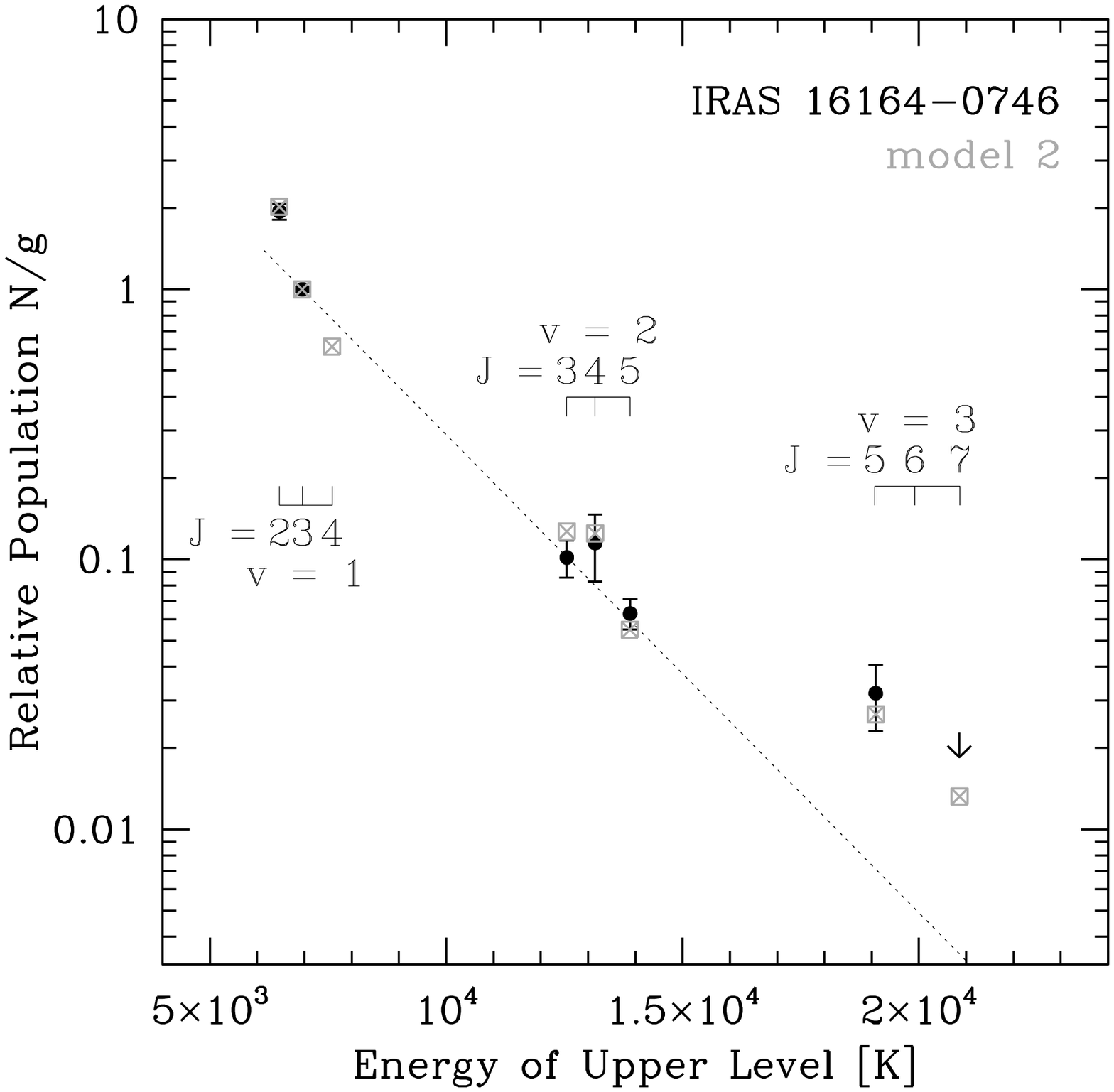,width=8cm}\hspace{5mm}\psfig{file=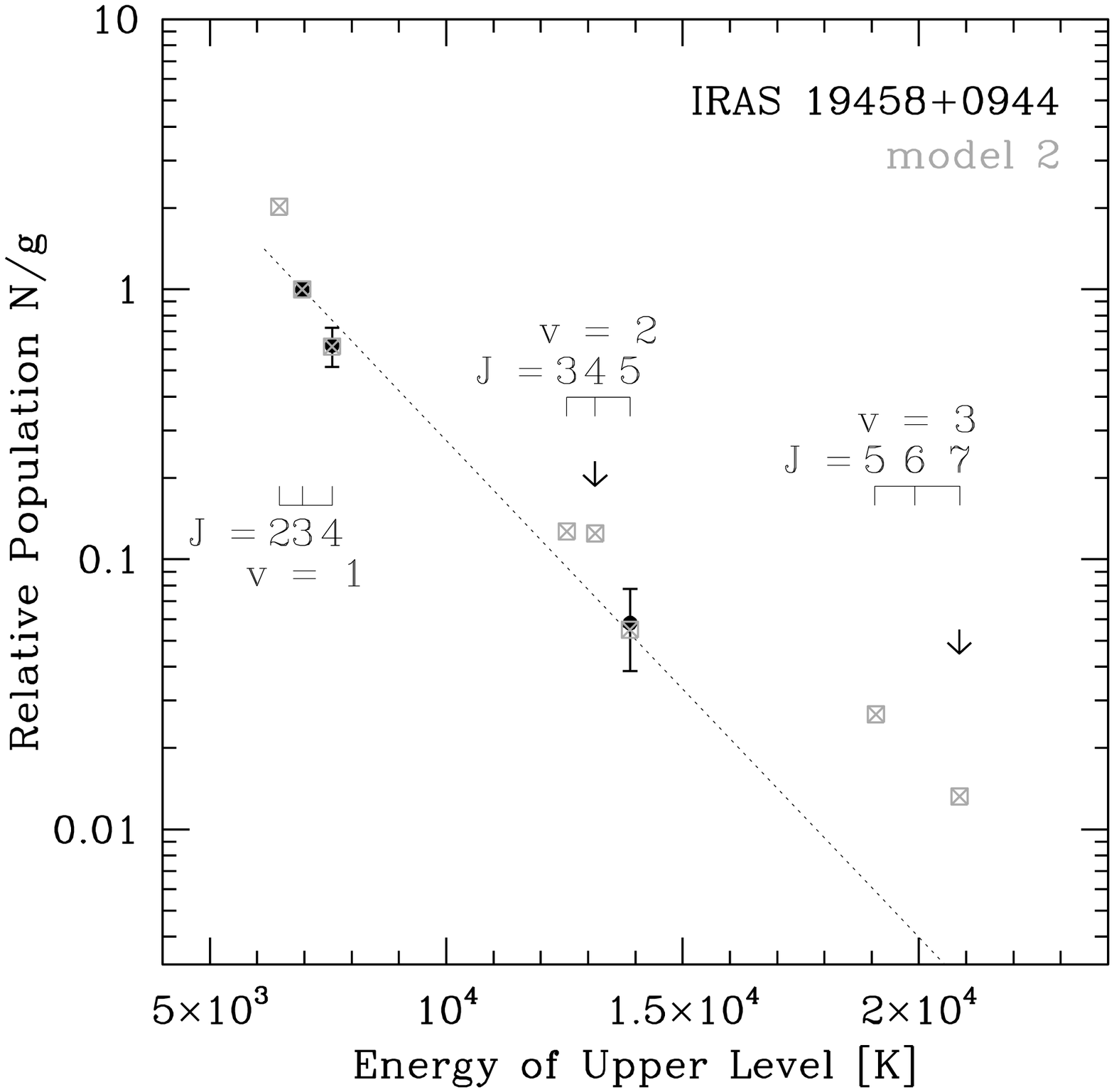,width=8cm}}
\vspace{5mm}
\centerline{\psfig{file=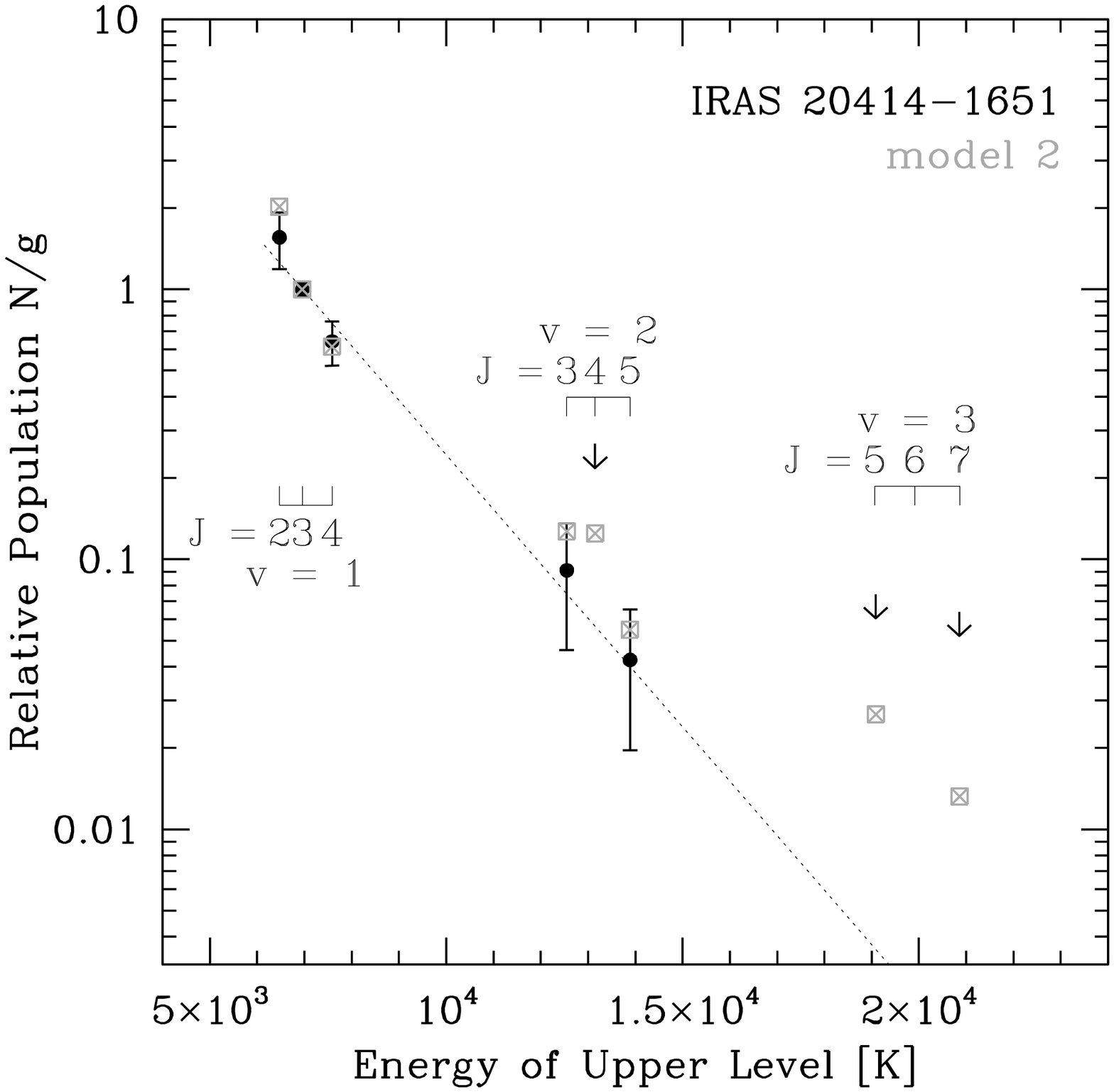,width=8cm}\hspace{5mm}\hspace{8cm}}
%\caption{As for Fig.~\ref{fig:pop}.}
%\label{fig:popb}
\end{figure}

%----------------------------------------------------------------------

\begin{figure}
\centerline{\psfig{file=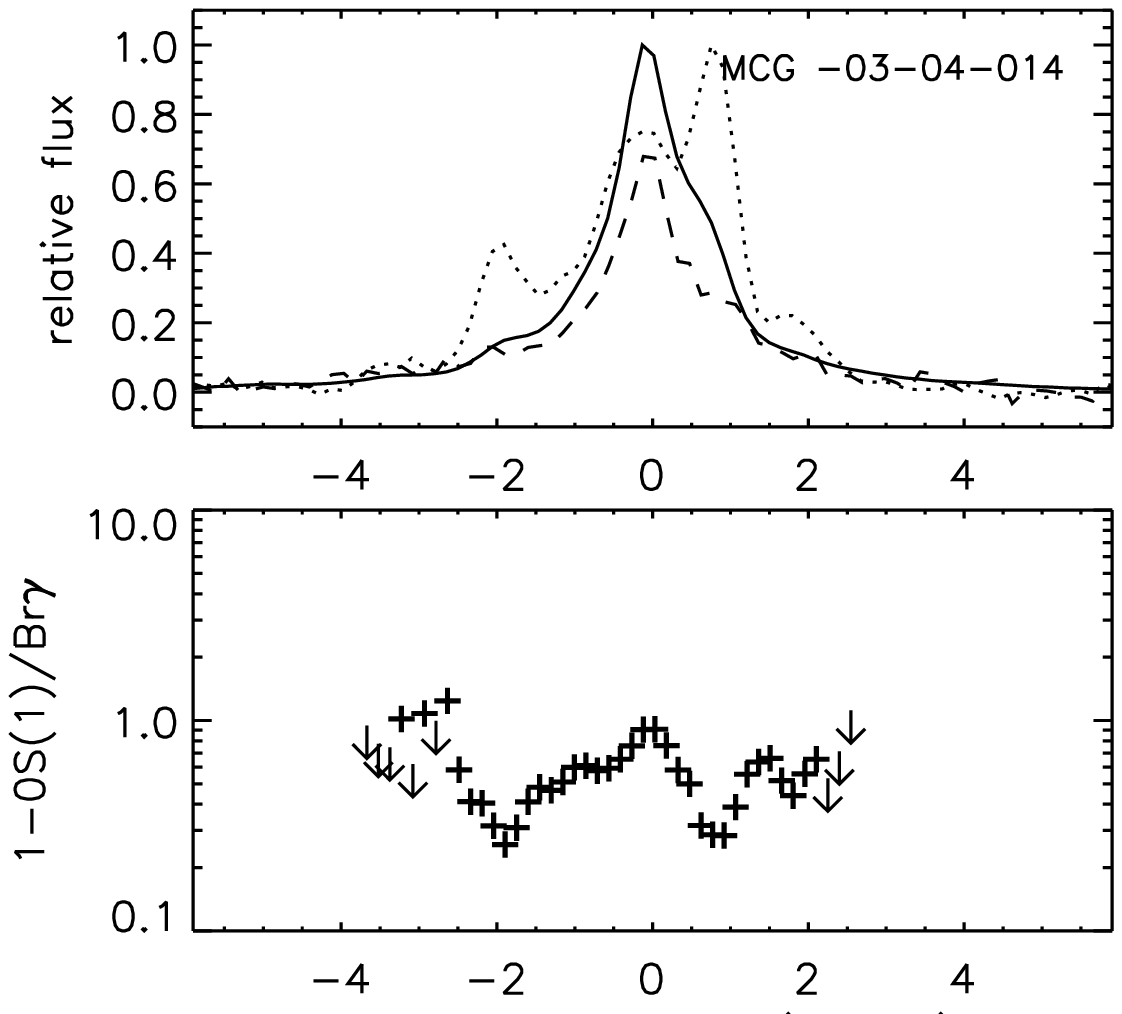,width=5.5cm}\psfig{file=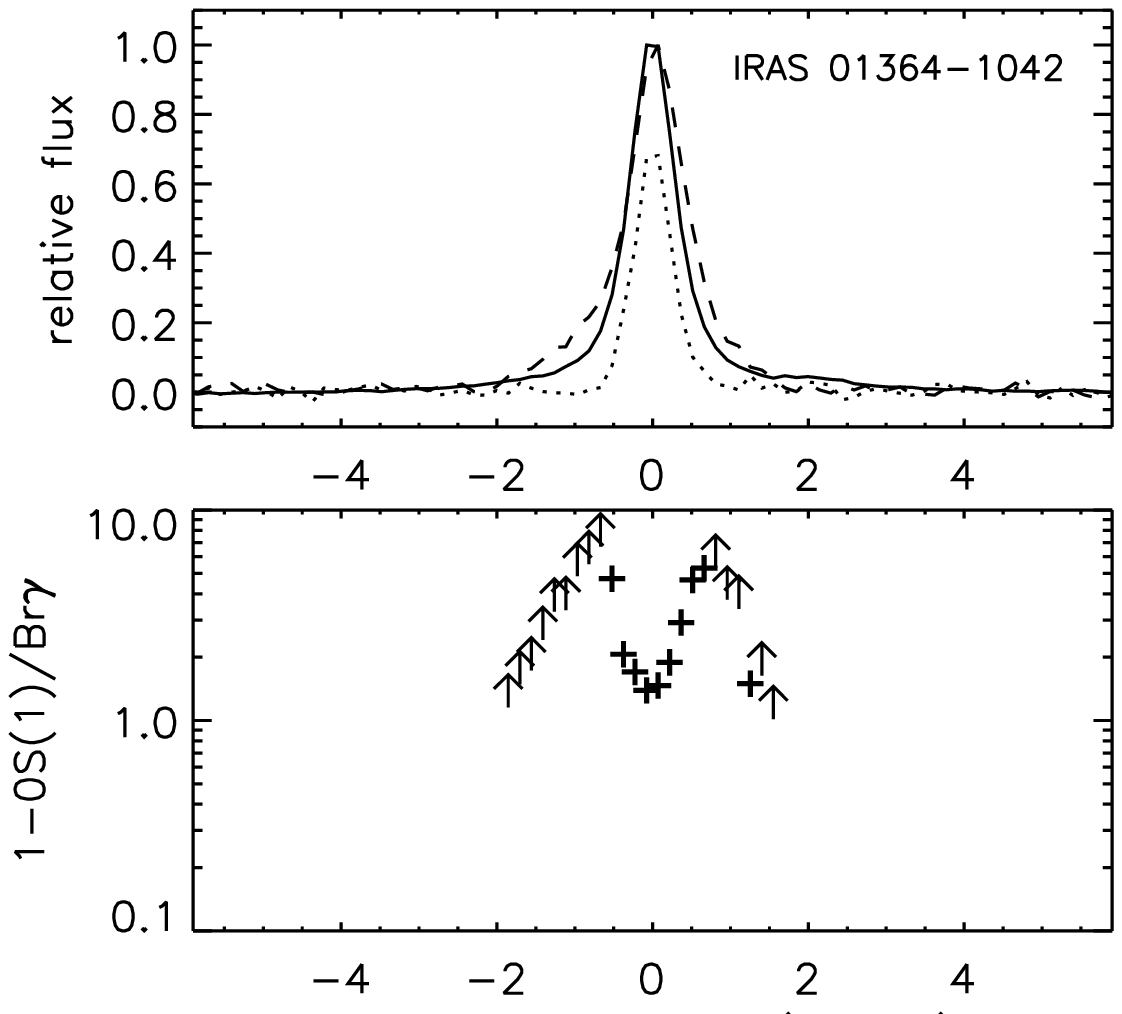,width=5.5cm}\psfig{file=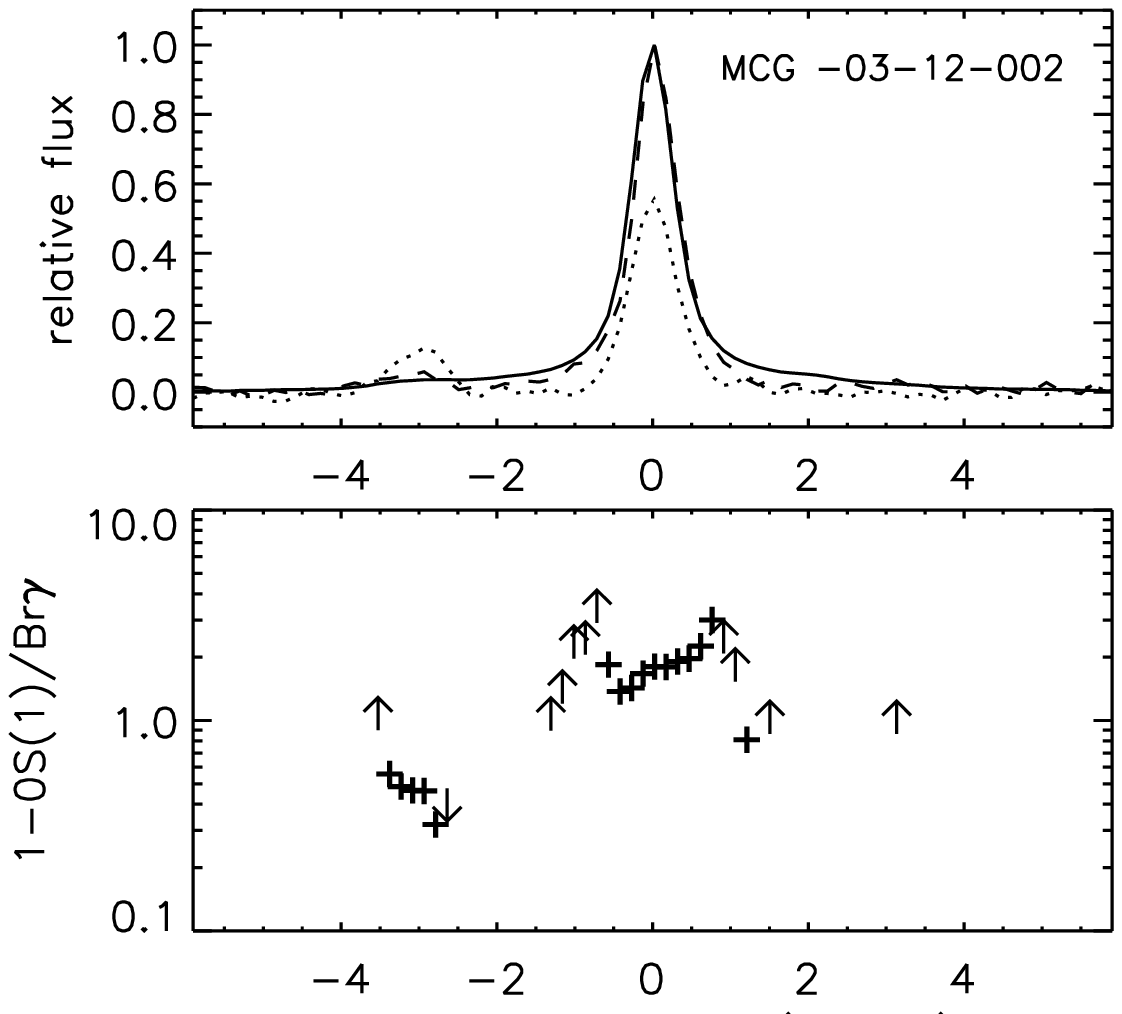,width=5.5cm}}
\vspace{7mm}
\centerline{\psfig{file=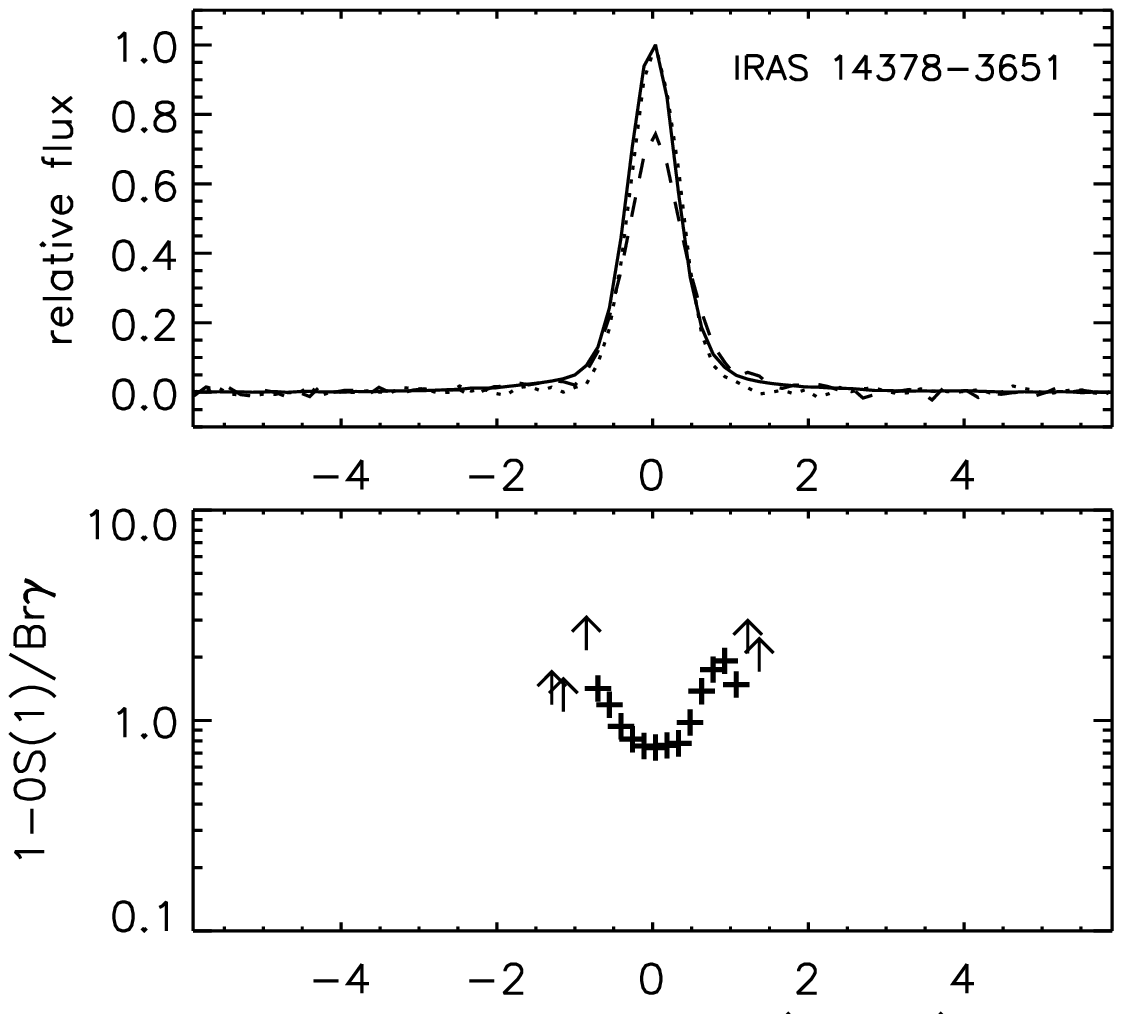,width=5.5cm}\psfig{file=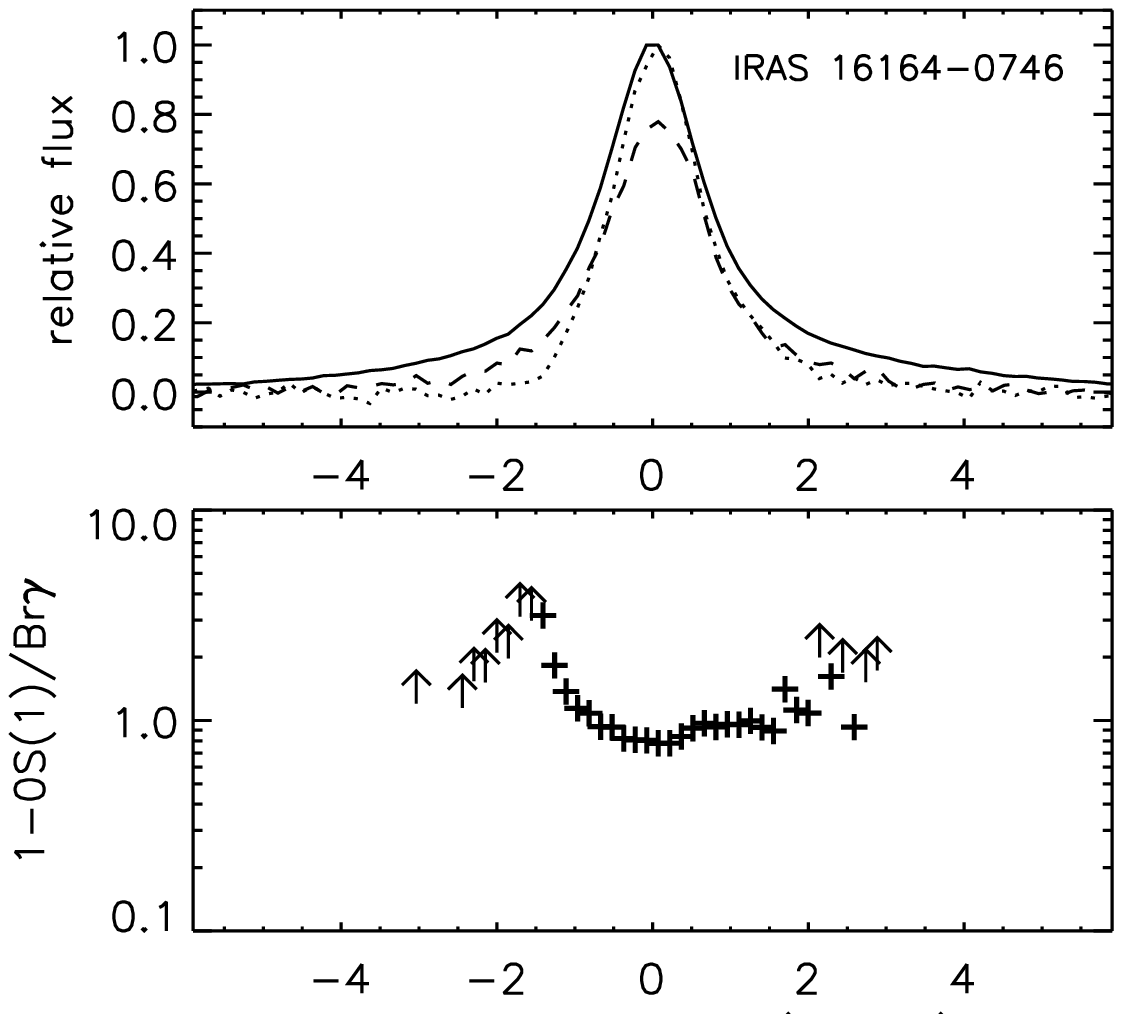,width=5.5cm}\psfig{file=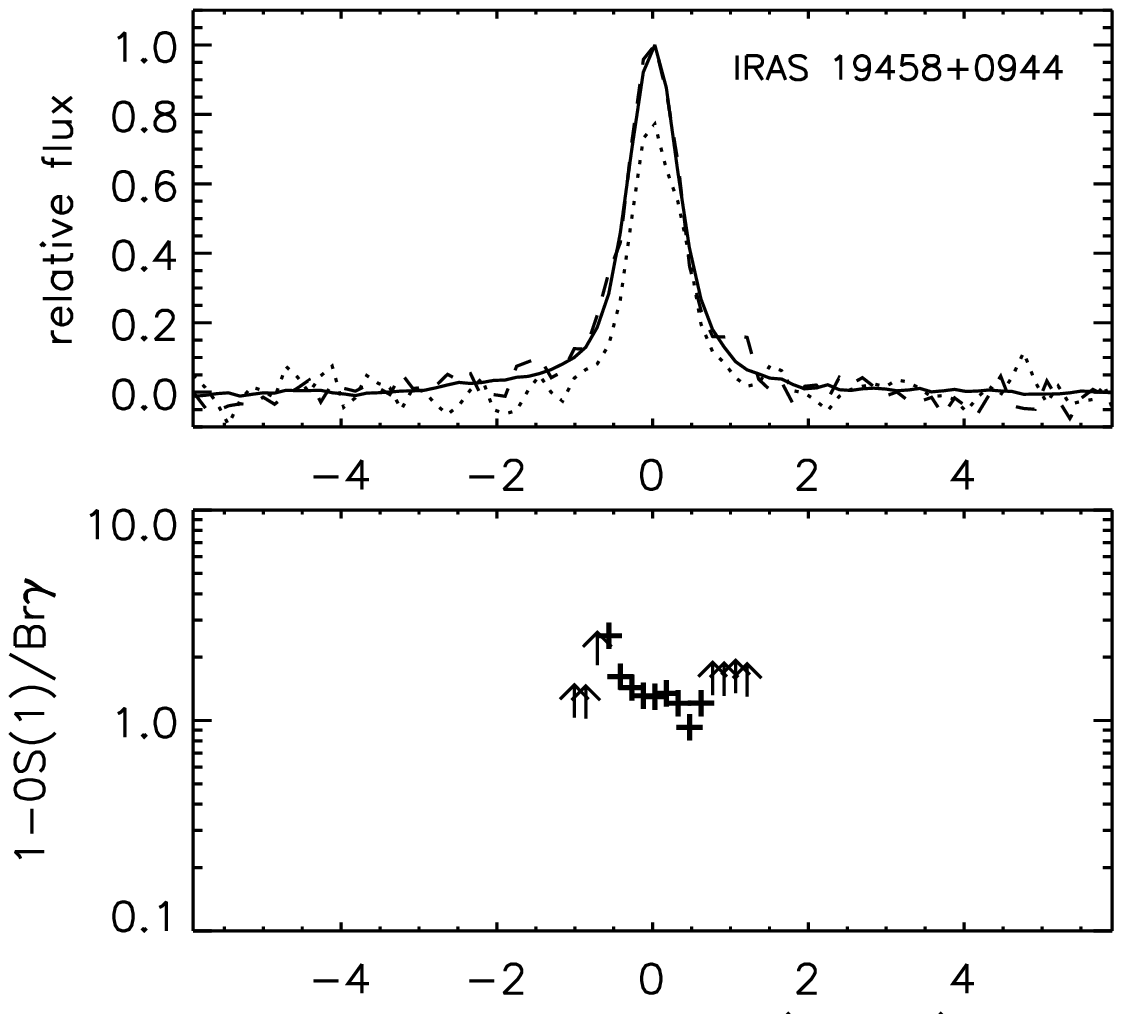,width=5.5cm}}
\vspace{7mm}
\centerline{\psfig{file=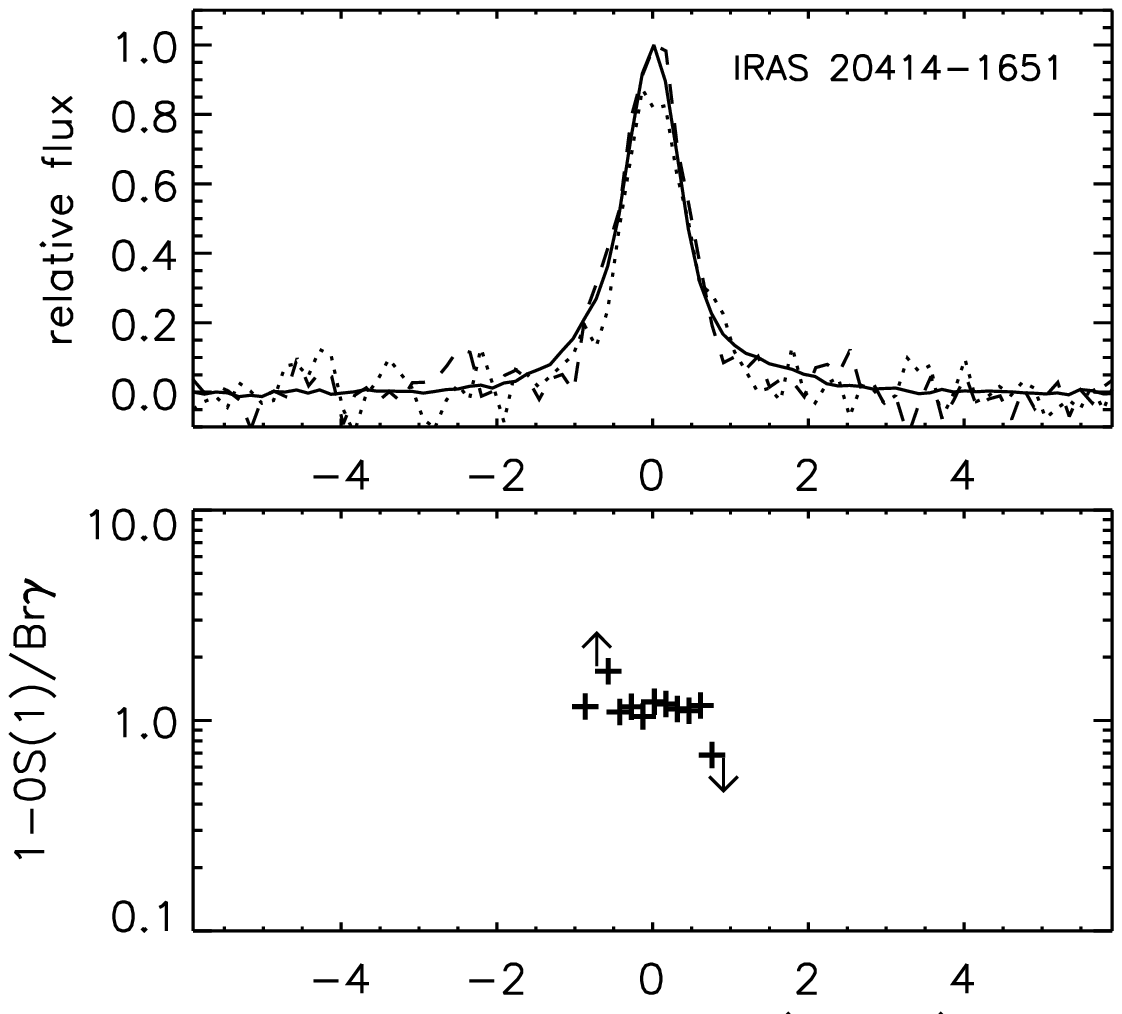,width=5.5cm}\hspace{11cm}}
\caption{Upper: Normalised spatial distributions of the continuum flux
density (solid line), the Br$\gamma$ flux (dotted line), and the
1-0\,S(1) flux (dashed line).
A single normalisation was applied to the Br$\gamma$ and 1-0\,S(1)
fluxes.
Lower: 1-0\,S(1)/Br$\gamma$ ratio (arrows denote $3\sigma$ limits).
}
\label{fig:spat}
\end{figure}

%----------------------------------------------------------------------

\begin{figure}
\centerline{\psfig{file=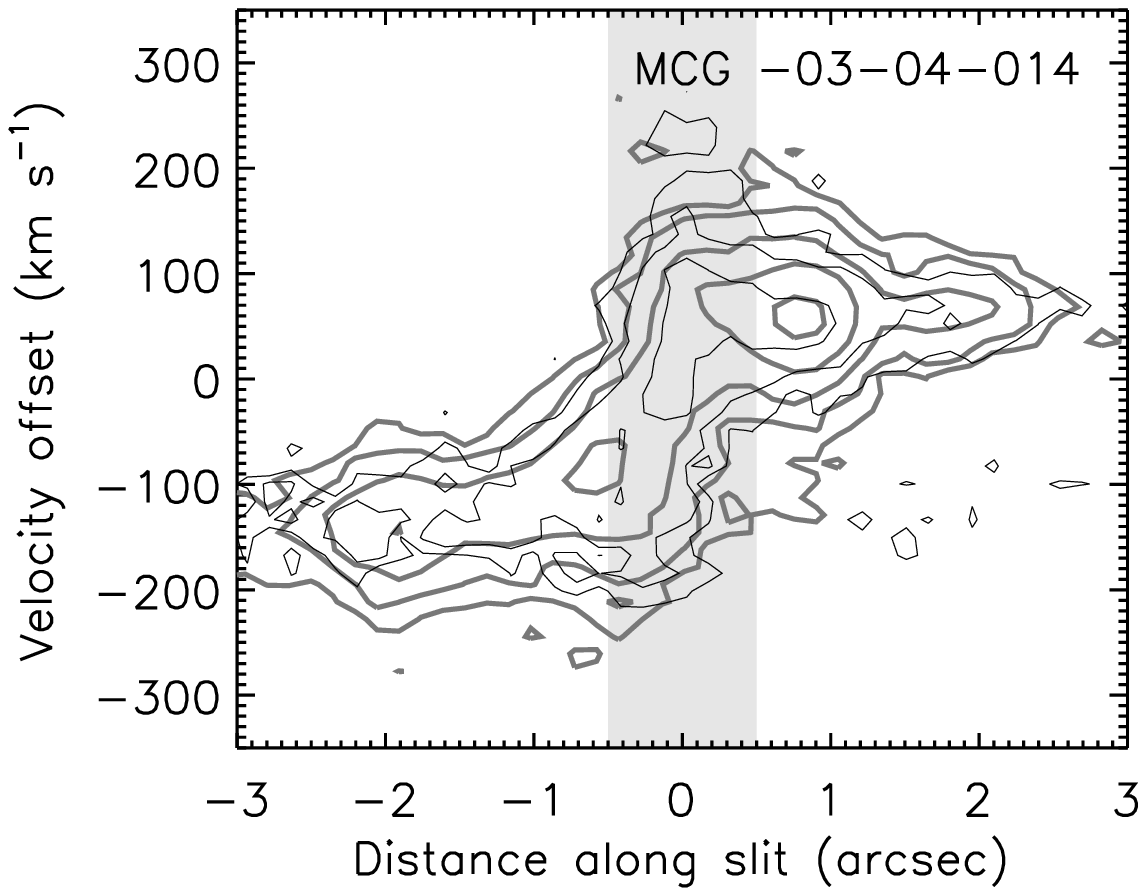,width=6cm}\psfig{file=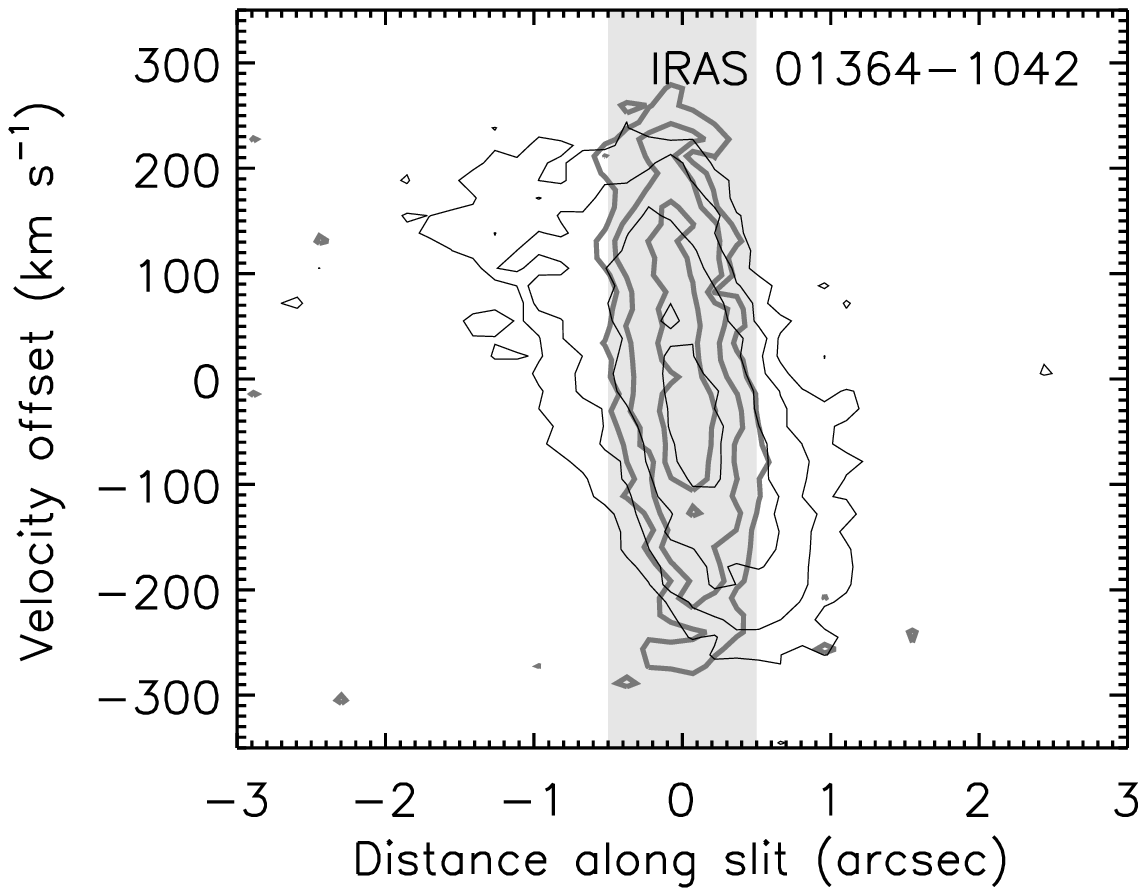,width=6cm}\psfig{file=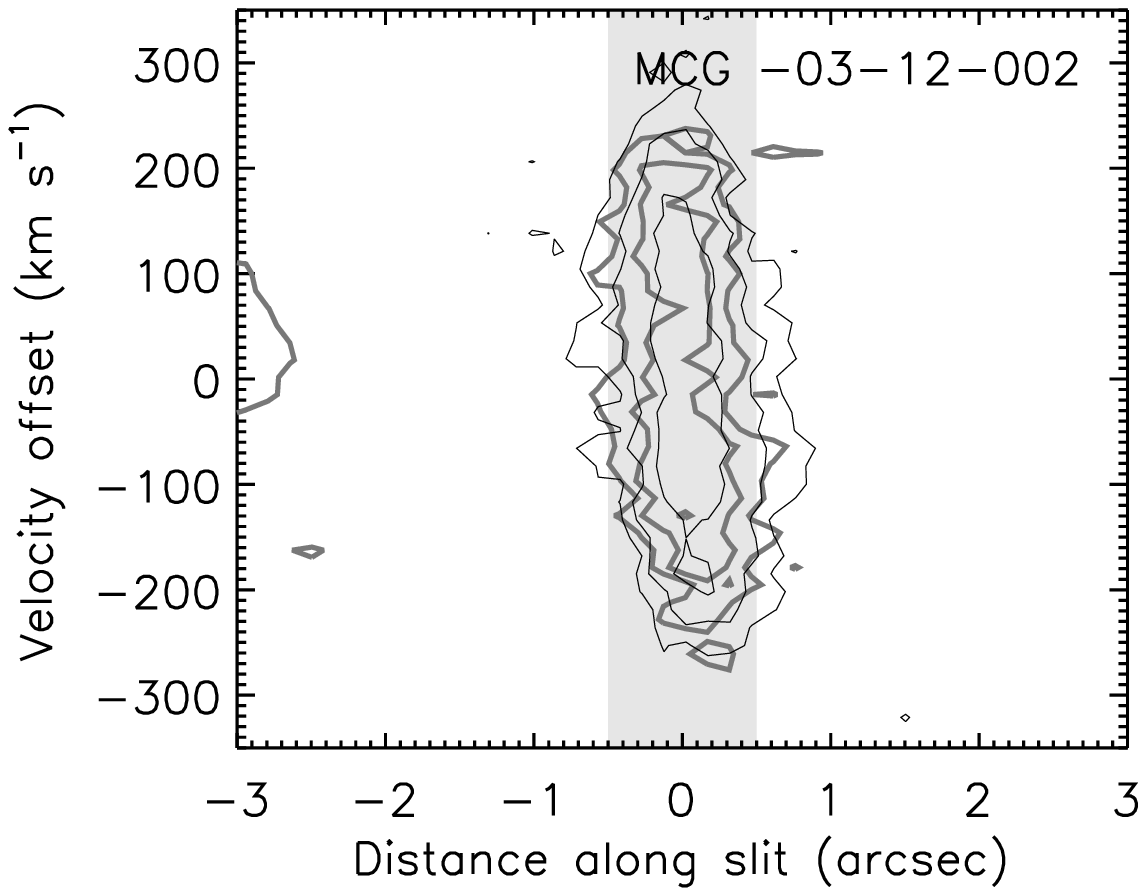,width=6cm}}
\vspace{1mm}
\centerline{\psfig{file=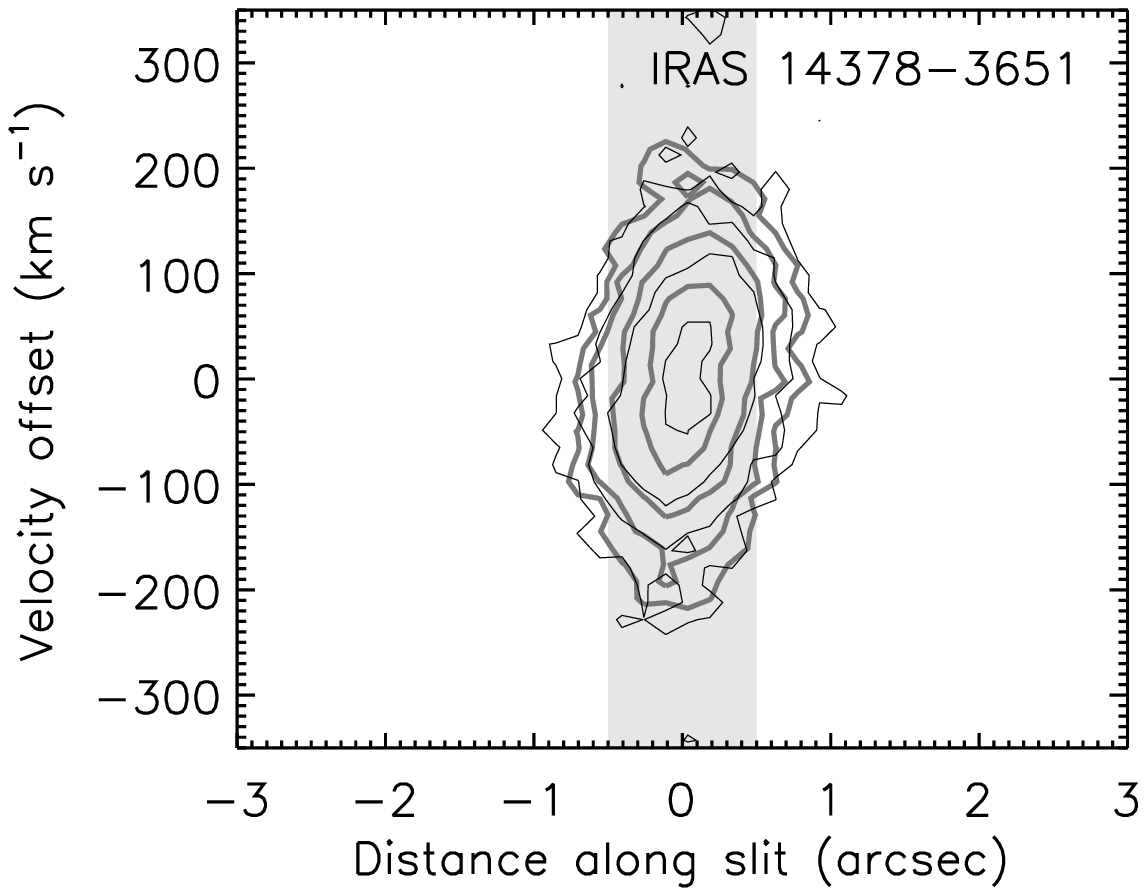,width=6cm}\psfig{file=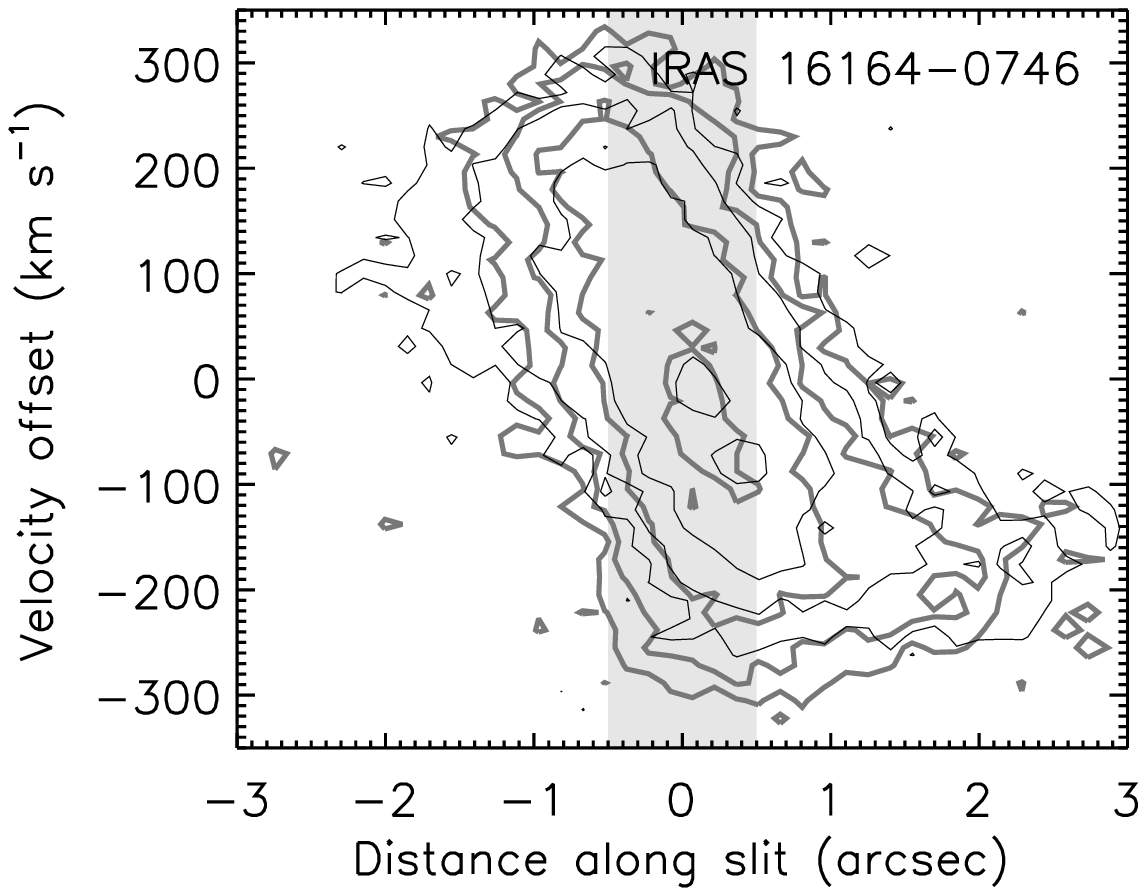,width=6cm}\psfig{file=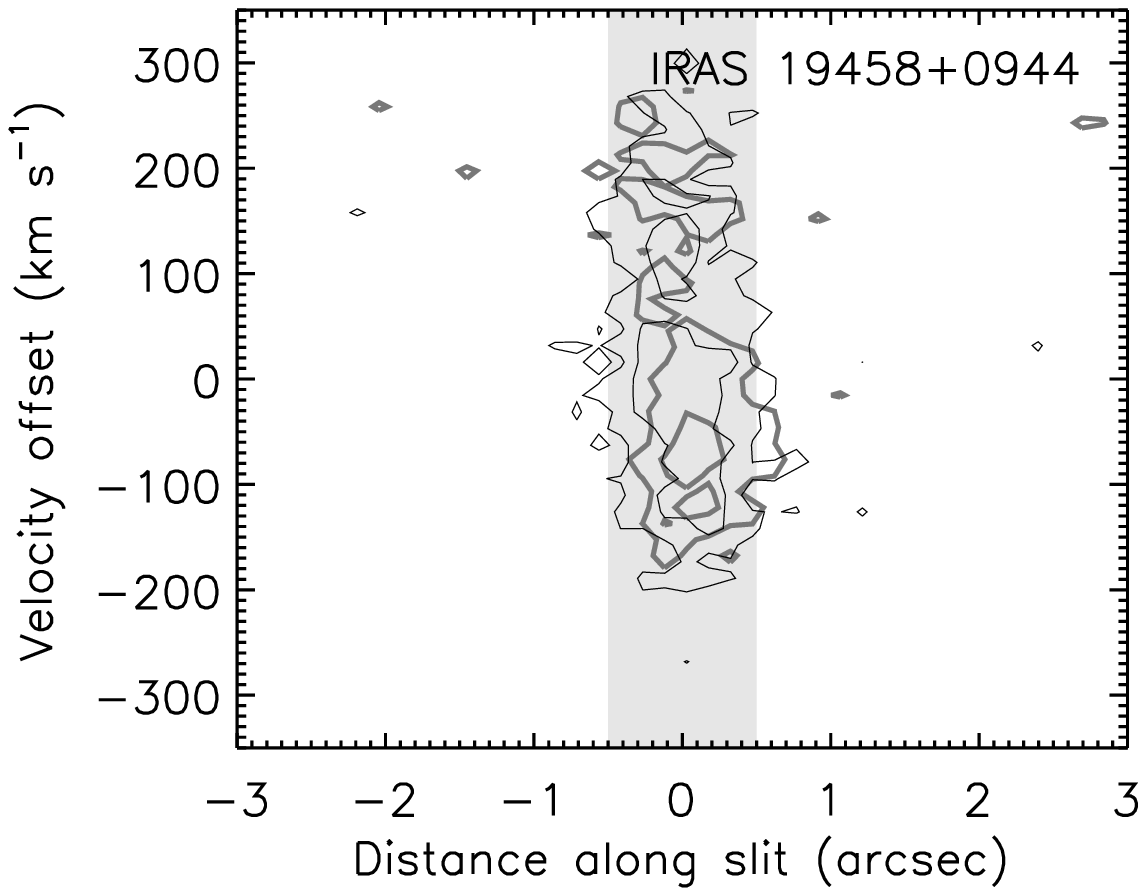,width=6cm}}
\vspace{1mm}
\centerline{\psfig{file=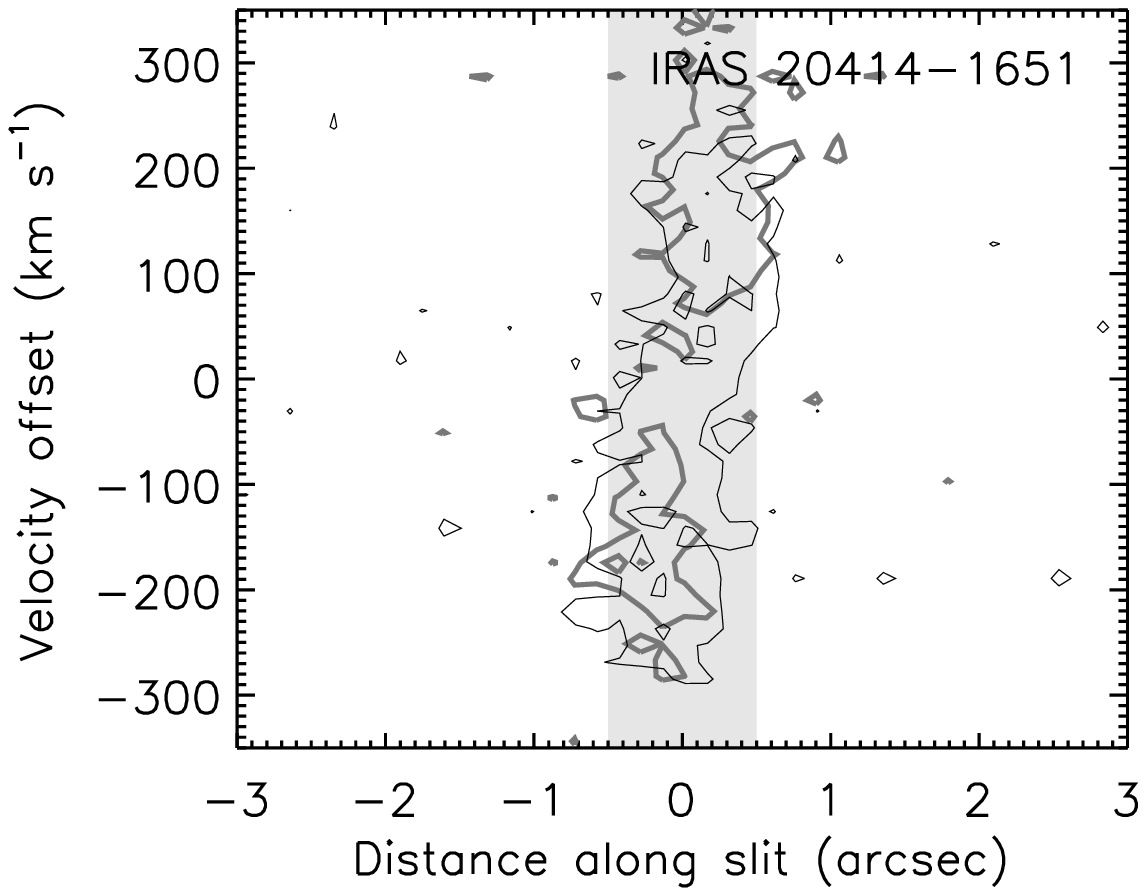,width=6cm}\hspace{12cm}}
\caption{
Position-velocity diagrams.
The thick grey line is the Br$\gamma$, while the thin black line is
for 1-0\,S(1).
The pale vertical stripe shows the region over which the spectra were
extracted.}
\label{fig:vel}
\end{figure}

%----------------------------------------------------------------------

\end{document}